\documentclass{iopart}
\newcommand{\g}{\mathcal{G}}
\newtheorem{example}{Example}
\newtheorem{proposition}{Proposition}

\newtheorem{definition}{Definition}
\usepackage{iopams}
\usepackage[all]{xy}
\usepackage{multirow}

\begin{document}

\title[Contraction of representations]{Strong contraction of the representations of the three dimensional Lie algebras}

\author{E M Subag$^{1}$, E M Baruch$^{1}$, J L Birman$^{2}$ and A Mann$^{3}$}

\address{$^1$ Department of Mathematics, Technion-Israel Institute of Technology, Haifa 32000, Israel\newline
  $^2$ Department of Physics, The City College of the City University of New York, New York N.Y.10031, USA\newline
 $^3$ Department of Physics, Technion-Israel Institute of Technology, Haifa 32000, Israel}
\ead{esubag@tx.technion.ac.il, embaruch@math.technion.ac.il, birman@sci.ccny.cuny.edu and ady@physics.technion.ac.il}

\begin{abstract}
For any \.{I}n\"{o}n\"{u}-Wigner contraction of a three dimensional Lie algebra we construct the corresponding contractions of representations. Our method is quite canonical in the sense that in all cases we  deal with realizations of the representations on some spaces of functions;  we contract the differential operators on those spaces along with the representation spaces themselves by taking certain pointwise limit of functions. We call such contractions strong contractions. We show that this pointwise limit gives rise to a direct limit space.  Many of these contractions are new and in other examples we give a different proof.
\end{abstract}


\section{Introduction}
In this paper we obtain contractions of Lie algebra representations for all the proper \cite{Wei3} \.{I}n\"{o}n\"{u}-Wigner contractions (IW-contractions) \cite{IW1,IW2} of three dimensional Lie algebras. More precisely, if a three dimensional Lie algebra $\g_0$ is a contraction of $\g$ then for any irreducible representation $\eta$ of $\g_0$ that comes from some unitary representation of the corresponding Lie group we find a family of representations of $\g$ that contract to $\eta$.  We do not consider the rather  trivial case of one dimensional representations. Some contractions of representations of three dimensional Lie algebras were obtained  in  e.g., \cite{IW1,Wei2,CeTa1,Bar,Bar2}. More examples can be found by e.g., \cite{Gil,CeTa3,Wei44,Faw,Tal}.
We present a systematic and unified  approach to contraction of the representations of the three dimensional Lie algebras.  Some of the contractions that we obtain are new and in the other cases we present a new proof.
The type of convergence that we obtain is as follows. We start with  a sequence of functions  $\left\{{f_n(x)}\right\}_{n=0}^{\infty}$ that converges   to $f(x)$ pointwise and in $L^2$ norm.  For a sequence of differential operators $\left\{{\rho_n}\right\}_{n=0}^{\infty}$  we get
\begin{eqnarray}
&&\rho_n(f_n(x)) \; \; \mbox{converges to} \;\rho_{\infty}(f(x))
\label{Eq1}
\end{eqnarray}
where $\rho_{\infty}$ is some differential operator on our space and the convergence is pointwise  and in  $L^2$ norm. We call such contractions strong contractions of representations.\newline

It seems that the strong contraction procedure, as it is based on representations that are realized  on some functions spaces, possess a geometrical interpretation. It is possible to think of contraction as a process of flattening a curved space.
On the level of Lie algebras this is reflected by the fact that after contraction we
get a more abelian Lie algebra. When the group or the symmetric space is contracted
we get a flatter space (See  \cite{Gil,MickN,Ball}).  On the other hand, when contracting the representation spaces using
limits of matrix elements it is very difficult to see the flattening procedure or how the limiting
parameter makes the spaces more flat. Contracting using differential operators reveals the flattening procedure.
To illustrate this point we consider the famous contraction of \.{I}n\"{o}n\"{u} and Wigner \cite{IW1}, contracting  representations of $su(2)$  into a
representation of $iso(2)$. The contraction process takes a series of finite dimensional representations of increasing dimension and "contracts" them into an infinite dimensional representation of $iso(2)$. It is not clear why the resulting representation is more "flat" than the series from which it is obtained. On the other hand, consider the following "strong contraction" of the same representations.
We realize the finite dimensional representation of dimension $2l+1$ as the space of spherical functions of degree $l$ spanned by the spherical harmonics
$Y_l^m$. These can be thought of as functions of the angular coordinates $(\theta,\varphi)$. We can
identify a point on the sphere with a point on the disk with radius $\theta$ and angle $\varphi$. We dilate the disk by $l$ by multiplying the $\theta$ coordinate by $l$ to obtain the coordinates $(l \theta,\varphi)$. Applying the same transformation to the $Y_l^m$ we get functions on a disk of radius $l\pi$ (or a sphere of radius $l$). Letting $l$ go to infinity we obtain a limit of functions on increasing size disks or spheres thus obtaining a function on a flat space (the Euclidean plane). We show (subsection \ref{C7}) that such a
limit exists and induces the same contraction considered by \.{I}n\"{o}n\"{u} and Wigner. Our construction illustrates the geometric effect of the parameter $l$ as a contraction parameter.\newline

The strong contraction approach allowed us to obtain many new contractions. For a list of those see \ref{se}.\newline

  Our paper is divided as follows: In the next  section  we recall the definition for contraction of Lie
algebras. In section \ref{sec3} we give a definition for strong contraction of Lie algebra representations. In section \ref{sec4} we list all the three dimensional real Lie algebras and their  proper IW-contractions. In section \ref{sec5} we give explicitly all contractions of Lie algebra representations for each proper IW-contraction of a real three dimensional Lie algebra.
In the appendix we review briefly the notion of direct limit in the category of inner product spaces and give a definition for contraction of Lie algebra representations in terms of direct limit. We show that a strong contraction for which we also have a compatible family of bases gives a contraction by means of convergence of  matrix elements.
\section{Preliminaries}
Roughly speaking contraction of a Lie algebra/group is the process of obtaining one Lie algebra/group as some kind of limit from another Lie algebra/group.
The first to investigate such kind of limits was Segal \cite{Segal} in 1951. He was motivated by the fact that quite often one physical theory approximates or tends under some limit to another physical theory.
His idea was that if two theories (e.g., relativistic and classical mechanics) are related by a limiting process, then the associated Lie groups/algebras (e.g., Poincar\'{e} and Galilei groups) should also be related by some limiting process. This idea was studied further in 1953 by \.{I}n\"{o}n\"{u} and Wigner \cite{IW1,IW2} who introduced the so-called  \.{I}n\"{o}n\"{u}-Wigner contractions (IW-contractions). In 1961 Saletan \cite{Sal} defined a contraction of Lie algebras and the IW-contractions are special cases of his definition. We now give
the formal definition for  contraction of Lie algebras using notations that are similar to those
of Weimar-Woods \cite{Wei3}.
\begin{definition}
Let $\g=\left(U, [\_,\_] \right)$  be a  Lie algebra with an underlying vector space $U$ and a Lie bracket $[\_,\_]$. For every $\epsilon \in (0,1]$ let $t_{\epsilon}:U\longrightarrow U$ be an invertible linear  transformation. If for every $X,Y \in U$
\begin{eqnarray}
\lim_{\epsilon \longrightarrow 0^{+}}t^{-1}_{\epsilon}([t_{\epsilon}(X),t_{\epsilon}(Y)])
\label{eq:1}
\end{eqnarray}
exists we will denote it by $[X,Y]_0$. In this case $\g_0=\left(U,[\_,\_]_0 \right)$ is a Lie algebra
and we call $\g_0$ the contraction of $\g$ by  $t_{\epsilon}$ and we write $\mathcal{G}\stackrel{t(\epsilon)}{\rightarrow} \mathcal{G}_0$.
\end{definition}
If there are bases for $U$ such that the matrix that represents $t_{\epsilon}$ relative to those bases is of the form $diag(1,1...1,\epsilon,\epsilon,...\epsilon)$ then the contraction is called a simple IW-contraction or just an IW-contraction.\newline
There is an analogous definition  for the case that  the limit ($\ref{eq:1}$) is meaningful only on a sequence $\left\{\epsilon_{n}\right\}_{n=0}^{\infty}$ which converges to zero when $n \longrightarrow \infty$.
\begin{definition}
Let $\g=\left(U, [\_,\_] \right)$  be a  Lie algebra  with an underlying vector space $U$ and a Lie bracket $[\_,\_]$. For every $n \in \mathbb{N}$ let $t_n:U\longrightarrow U$ be an invertible linear  transformation. If for every $X,Y \in U$
\begin{eqnarray}
\lim_{n \longrightarrow \infty}t^{-1}_{n}([t_{n}(X),t_{n}(Y)])
\label{eq:2}
\end{eqnarray}
exists we will denote it by $[X,Y]_{\infty}$. In this case $\g_{\infty}=\left(U,[\_,\_]_{\infty} \right)$ is a Lie algebra
and we call $\g_{\infty}$ the sequential contraction of $\g$ by  $t_n$ and we write $\mathcal{G}\stackrel{t_n}{\rightarrow} \mathcal{G}_{\infty}$. We sometimes call a sequential contraction just a contraction. By abuse of notation we sometimes denote $\g_{\infty}$ also  by $\g_0$.
\end{definition}
Specific examples of contractions of Lie algebras can be found for example in  \cite{IW1,Gil,Sal,Wei1}. For generalizations of IW-contractions see for example \cite{Wei00,Mon}.
\section{Contraction of representations as pointwise limit.}
\label{sec3}
In the discrete case, i.e., when $\mathcal{G}\stackrel{t_n}{\rightarrow} \mathcal{G}_{\infty}$ and we consider a sequence of representations of $\g$, our setting is as follows. For each $n\in \mathbb{N}$ let $X_n$ be a set such that $X_{n+1}\supseteq X_n$. Let $V_n$ be an  inner product subspace  of $L^2(X_n,d\mu_n)$ where $\mu_n$ is a (positive) measure on $X_n$. Suppose that for every $m,n \in \mathbb{N}$ such that $n\geq m$ we have an isometric embedding $\varphi_{mn}:V_m \longrightarrow V_n$ and assume that the embeddings are compatible in the following sense; $\varphi_{nk}\circ\varphi_{mn}=\varphi_{mk}$ for any $m,n,k \in \mathbb{N}$ such that $m\leq n\leq k$. Let $X=\bigcup_{n=1}^{\infty}X_n$, and let $V$ be an inner product subspace of $L^2(X,d\mu)$ where $\mu$ is a (positive) measure on $X$.

\begin{definition}
\label{def3}
In the above setting suppose $\g\stackrel{t_n}{\rightarrow} \g_{\infty}$ and  $\left\{(\rho_{n},V_{n})\right\}_{n \in  \mathbb{N}}$ is a sequence of representations of $\g$. A representation $\eta:\g_{\infty}\longrightarrow gl(V)$ is called the strong  contraction of the sequence of representations $\left\{(\rho_{n},V_{n})\right\}_{n \in  \mathbb{N}}$ with respect to
the contraction $\g\stackrel{t_n}{\rightarrow} \g_{\infty}$, and we denote it by $\rho_n\stackrel{t_n}{\rightarrow} \eta$, if the following conditions hold:
\begin{enumerate}
	\item For any $m_0\in \mathbb{N}$ any $f \in V_{m_0}$ and any $x\in X$, the pointwise limit, $\lim_{n \longrightarrow \infty} \varphi_{m_0n}(f)(x)$ exists and defines a function in $V$ which we denote by $L(f)$.
	\item For every $F\in V$ there exists a function $f$ in some $V_{m_0}$ such that $L(f)=F$.
	\item $L$ preserves the inner products in the following sense: for every $f,g\in V_{m_0}$  we have $\left\langle L(f),L(g)\right\rangle=\left\langle f,g\right\rangle_{m_0}$.
 	\item For every $f\in V_{m_0}$, $Y\in U$ the sequence $L(\rho_n(t_{\epsilon_n}(Y))\varphi_{{m_0}n}(f))$ converge pointwise and in norm to $\eta(Y)L(f)$ i.e., for every  $x\in X$   $$ \lim_{n\longrightarrow \infty}L(\rho_n(t_{\epsilon_n}(Y))\varphi_{{m_0}n}(f))(x)=\eta(Y)L(f)(x)$$ and $$ \lim_{n\longrightarrow \infty}\|L(\rho_n(t_{\epsilon_n}(Y))\varphi_{{m_0}n}(f))-\eta(Y)L(f)\|=0$$
\end{enumerate}
\end{definition}
\textbf{Remark 1.}
\begin{itemize}
\item \textit{Since all the $\varphi_{mn}$ are compatible isometric embeddings we can rewrite  condition  $(iii)$ in the following way: for every $f\in V_{m_0}$, $g\in V_{n_0}$  we have $\left\langle L(f),L(g)\right\rangle=\lim_{k\longrightarrow \infty}\left\langle \varphi_{m_0k}(f),\varphi_{n_0k}(g)\right\rangle_k$. In fact this is a limit of a constant sequence since for any $k\geq M\equiv \max{\left\{m_0,n_0\right\}}$, $\left\langle \varphi_{m_0k}(f),\varphi_{n_0k}(g)\right\rangle_k = \left\langle \varphi_{Mk}\circ\varphi_{m_0M}(f),\varphi_{Mk}\circ\varphi_{n_0M}(g)\right\rangle_k= \left\langle \varphi_{m_0M}(f),\varphi_{n_0M}(g)\right\rangle_M$.}
\item \textit{One can easily show that under the assumptions of the above definition the inner product space $V$ is the  direct limit (see appendix) of the inner product spaces $\left\{V_n\right\}_{n \in \mathbb{N}}$.}
\item \textit{In all the cases that we will consider $V_n$ and $V$ are  dense subspaces of $L^2(X_n,d\mu_n)$ and $L^2(X,d\mu)$, respectively.}\newline
\end{itemize}

In the continuous case, i.e., when the contraction parameter varies in $(0,1]$  and we consider a family of representations of $\g$, our setting is as follows. Let $\left\{X_{\epsilon}\right\}_{\epsilon \in (0,1]}$ be a family of sets
such that for $\epsilon_2\geq \epsilon_1$ in $(0,1]$,  $X_{\epsilon_2}\subseteq X_{\epsilon_1}$. As in the discrete case we have a family of inner product spaces $V_{\epsilon}\subseteq L^2(X_{\epsilon},d\mu_{\epsilon})$ and whenever $\epsilon_2,\epsilon_1\in (0,1]$ satisfy $\epsilon_2\geq \epsilon_1$ we have an    isometric embeddings $\varphi_{\epsilon_2,\epsilon_1}:V_{\epsilon_2}\longrightarrow V_{\epsilon_1}$ and assume that these embeddings are compatible in the following sense; $\varphi_{\epsilon_n \epsilon_k}\circ\varphi_{\epsilon_m \epsilon_n}=\varphi_{\epsilon_m \epsilon_k}$ for any $\epsilon_m,\epsilon_n,\epsilon_k \in (0,1]$ such that $\epsilon_m \geq \epsilon_n\geq \epsilon_k$. We set  $X=\bigcup_{\epsilon \in (0,1]}X_{\epsilon}$, and let $V$ be an inner product subspace of $L^2(X,d\mu)$ where $\mu$ is some (positive) measure on $X$. Suppose $\g\stackrel{t(\epsilon)}{\rightarrow} \g_{0}$ and $\left\{(\rho_{\epsilon},V_{\epsilon})\right\}_{\epsilon \in  (0,1]}$ is a family of representations of $\g$. A representation $\eta:\g_0\longrightarrow gl(V)$ is called the strong contraction of the family of representations $\left\{(\rho_{\epsilon},V_{\epsilon})\right\}_{\epsilon \in  (0,1]}$ with respect to
the contraction $\g\stackrel{t(\epsilon)}{\rightarrow} \g_{0}$ and we denote it by $\rho_{\epsilon}\stackrel{t(\epsilon)}{\rightarrow} \eta$ if analogous conditions to those in definition \ref{def3} hold where instead of taking the limit $n\longrightarrow \infty$  we take $\epsilon \longrightarrow 0^+$.
\newpage

\section{The IW-contractions of the real three dimensional Lie algebras}
\label{sec4}
In this section we list all the real three dimensional Lie algebras (up to an isomorphism) and following Conatser \cite{Conatser} we give the graph of their proper IW-contractions.

\begin{table}[h]
\caption{List of all the three dimensional real Lie algebras is given by specifying the non vanishing commutation relations between the elements of the basis $X_1,X_2, X_3$. }
\begin{tabular*}{\textwidth}{@{}l*{15}{@{\extracolsep{0pt plus12pt}}l}}
\br
Our notation&Conatser's notation&Commutation relations\\
\mr
\hline $\mathfrak{ab}$ & $C_1$ &  $[\_,\_]\equiv 0 $
\\ \hline $\mathfrak{h}$ (Heisenberg)& $C_2$ &  $[X_3,X_2]=X_1 $
\\ \hline $\mathfrak{ea}$ & $C_3$ &  $[X_3,X_1]=X_1 $
\\ \hline $\mathfrak{g}(\lambda)$, $\lambda \in \mathbb{R}^{\ast}$ & $C_4(\lambda)$ &  $[X_3,X_1]=X_1$, $[X_3,X_2]=\lambda X_2 $
\\ \hline $\mathfrak{c}$ & $C_5$ &  $[X_3,X_1]=X_1 $, $[X_3,X_2]=X_1+X_2 $
\\ \hline
\multirow{2}{*}{$\mathfrak{l}(\lambda)$, $\lambda \in \mathbb{R}$} & $C_6(\lambda)$  & $[X_3,X_1]=X_2+\lambda X_1 $ \\ &  & $[X_3,X_2]=-X_1+\lambda X_2 $
\\ \hline
\multirow{3}{*}{$su(2)$} & $C_7$  & $[X_1,X_2]=X_3$ \\ &  & $[X_2,X_3]=X_1$
\\
&  & $[X_3,X_1]=X_2$
\\ \hline
\multirow{3}{*}{$sl_2(\mathbb{R})$} & $C_8$  & $[X_1,X_2]=X_3 $ \\ &  & $[X_2,X_3]=-X_1$
\\ &  & $[X_3,X_1]=-X_2 $
\\ \hline
\br
\end{tabular*}
\end{table}

\begin{figure}[h]
\caption{Graph of all the proper simple IW-contractions of real three dimensional Lie algebras. }
$\xymatrix{
& &su(2)\ar[drrr] & & & & sl_2(\mathbb{R})\ar[dddl]\ar[d]\ar[dl] \\
&\mathfrak{g}(1) &\mathfrak{c}\ar[l]\ar[rrrdd] &\mathfrak{g}(\lambda),\lambda \neq \pm1\ar[ddrr] &\mathfrak{l}(\lambda), \lambda \neq 0 \ar[ddr]  &iso(2)\ar[dd] &iso(1,1)\ar[ddl]  \\
& &  & & & &   \\
&  & &    & &\mathfrak{h}&\mathfrak{ea}\ar[l]   } $
\end{figure}
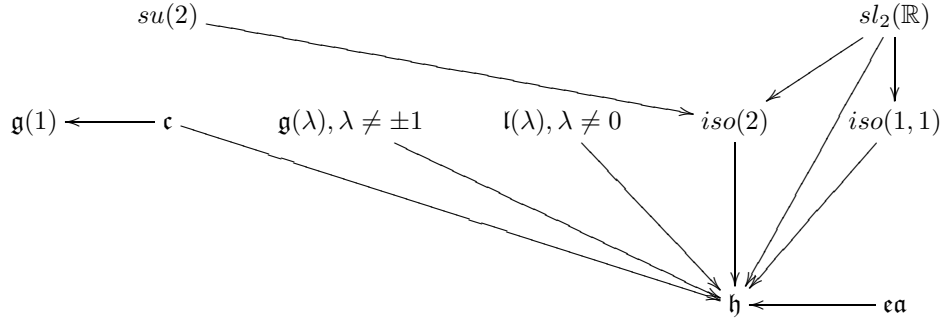

One should note that $\mathfrak{g}(-1)=iso(1,1)$ and that  $\mathfrak{l}(0)=iso(2)$. For more information on low-dimensional Lie algebras and their contractions see for example \cite{Wei2,Hud,Nes,Fla1,Fla2}.

\section{Contraction of Lie algebra representations for the real three dimensional Lie algebras}
\label{sec5}
In this section for every arrow in figure 1 we give explicitly a contraction of the corresponding Lie algebras. These contractions are essentially due to Conatser \cite{Conatser}.
Then for every proper IW-contraction of three dimensional real Lie algebras, $\g\stackrel{t(\epsilon)}{\rightarrow} \g_0$ we will specify all (except for the one dimensional ones)  the skew hermitian irreducible integrable representations (SHIIR) of $\g_0$.   Next for every such a representation $\eta$ of $\g_0$ we will find a family of representations of $\g$ that strongly contract to it. We always start with some family of SHIIR of $\g$  and then after intertwining it correctly in an $\epsilon$ dependent manner we obtain a suitable family that does strongly contract to $\eta$. \newline
An important ingredient in our procedure is choosing the suitable realization for each of the SHIIR that takes part in the above contractions. Many of these SHIIR are obtained naturally by Mackey's method \cite{Mack} for semidirect product groups, since every Lie algebra $\g_0$ which is an IW-contraction corresponds to a semidirect Lie group. In most of the other cases we take the appropriate realizations of the SHIRR from \cite{Vil}.\newline
In the rest of this section we will denote the interval $(0,1]$  by $I_1$.

\subsection{The contraction: $\mathfrak{ea}\longrightarrow \mathfrak{h}$}
\label{C1}Set $\g= \mathfrak{ea}$. The maps
\begin{eqnarray}
&& t_{\epsilon}:\mathfrak{ea}\longrightarrow \mathfrak{ea} \\ \nonumber
&& t_{\epsilon}(X_2^{\mathfrak{ea}})=\epsilon X_2^{\mathfrak{ea}} \\ \nonumber
&&  t_{\epsilon}(X_1^{\mathfrak{ea}}+X_2^{\mathfrak{ea}})=X_1^{\mathfrak{ea}}+X_2^{\mathfrak{ea}}  \\ \nonumber
&& t_{\epsilon}(X_3^{\mathfrak{ea}})=-\epsilon X_3^{\mathfrak{ea}}
\end{eqnarray}
realize the contraction of the Lie algebra and lead to the limit Lie algebra $\g_0$ with bracket given by $[X_3,X_1+X_2]_0=X_2$
which induce the isomorphism
\begin{eqnarray}
&& \psi:{\g}_0\longrightarrow \mathfrak{h} \\ \nonumber
&& \psi(X_2)= X_1^{\mathfrak{h}} \\ \nonumber
&& \psi(X_1+X_2)=X_2^{\mathfrak{h}}  \\ \nonumber
&& \psi(X_3)= X_3^{\mathfrak{h}}
\end{eqnarray}
Using the Mackey machine one can show  that by running over all $A\in \mathbb{R}^{*}$, the maps
\begin{eqnarray}
&& \eta^{\mathfrak{h}}_A:h \longrightarrow gl(L_c^{2,\infty}(\mathbb{R},dx)) \\ \nonumber
&& \eta^{\mathfrak{h}}_A(X_1^{\mathfrak{h}})=  iA  \\ \nonumber
&& \eta^{\mathfrak{h}}_A(X_2^{\mathfrak{h}})= iAx \\ \nonumber
&& \eta^{\mathfrak{h}}_A(X_3^{\mathfrak{h}})=\frac{d}{dx} \nonumber
\end{eqnarray}
exhaust all the SHIIR of $\mathfrak{h}$ where $L_c^{2,\infty}(\mathbb{R},dx)$ stands for the inner product space of compactly supported smooth functions on $\mathbb{R}$ that are square integrable relative to the Lebesgue measure. Using Mackey machine again it can be shown that for every $(a,b)\in \mathbb{R} \times \mathbb{R}^{*}$ the map
\begin{eqnarray}
&& \rho^{\mathfrak{ea}}_{(a,b)}:ea \longrightarrow gl(L^{2,\infty}_c(\mathbb{R}^+,\frac{dx}{x})) \\ \nonumber
&& \rho^{\mathfrak{ea}}_{(a,b)}(X_1^{\mathfrak{ea}})= ibx \\ \nonumber
&& \rho^{\mathfrak{ea}}_{(a,b)}(X_2^{\mathfrak{ea}})= ia   \\ \nonumber
&& \rho^{\mathfrak{ea}}_{(a,b)}(X_3^{\mathfrak{ea}})= x\frac{d}{dx}\nonumber
\end{eqnarray}
is a SHIIR where $L_c^{2,\infty}(\mathbb{R}^+,\frac{dx}{x})$ stands for the inner product space of compactly supported smooth functions on $\mathbb{R}^+$ that are square integrable relative to the measure $\frac{dx}{x}$.\newline

For every $\epsilon \in I_1$ let $V_{\epsilon}$ be  $L_c^{2,\infty}(\mathbb{R},dx)$. We define a function $\psi_{\epsilon}:\mathbb{R}\longrightarrow \mathbb{R}^+$ by $\psi_{\epsilon}(x)=e^{-\epsilon x}$.
We note that $P_{\epsilon}:L_c^{2,\infty}(\mathbb{R}^+,\frac{dx}{x})\longrightarrow V_{\epsilon}$ which is given by $P_{\epsilon}(f)=f \circ \psi_{\epsilon}$ is  an isomorphism. Its inverse $P_{\epsilon}^{-1}$ is given by  $P_{\epsilon}^{-1}(f)=f \circ \psi^{-1}_{\epsilon}$.
For each  $\epsilon \in I_1$  we intertwine  $\eta^{\mathfrak{ea}}_{(a,b)}$ with $P_{\epsilon}$ to get the equivalent  representation  $\rho^{\mathfrak{ea}}_{(a,b,\epsilon)}$  given by:
\begin{eqnarray}
&& \rho^{\mathfrak{ea}}_{(a,b,\epsilon)}:ea \longrightarrow gl(V_{\epsilon}) \\ \nonumber
&& \rho^{\mathfrak{ea}}_{(a,b,\epsilon)}(X_1^{\mathfrak{ea}})  =  P_{\epsilon} \circ \rho^{\mathfrak{ea}}_{(a,b)}(X_1^{\mathfrak{ea}}) \circ P^{-1}_{\epsilon} =ibe^{-\epsilon x}  \\ \nonumber
&& \rho^{\mathfrak{ea}}_{(a,b,\epsilon)}(X_2^{\mathfrak{ea}})=P_{\epsilon} \circ \rho^{\mathfrak{ea}}_{(a,b)}(X_2^{\mathfrak{ea}}) \circ P^{-1}_{\epsilon}=ia   \\ \nonumber
&& \rho^{\mathfrak{ea}}_{(a,b,\epsilon)}(X_3^{\mathfrak{ea}})= P_{\epsilon} \circ \rho^{\mathfrak{ea}}_{(a,b)}(X_3^{\mathfrak{ea}}) \circ P^{-1}_{\epsilon}=-\frac{1}{\epsilon}\frac{d}{dx} \nonumber
\end{eqnarray}
For $\epsilon_i, \epsilon_j\in I_1$ such that $\epsilon_j \leq \epsilon_i$ we define the linear isometry  $\varphi_{\epsilon_i,\epsilon_j}:V_{\epsilon_i}\longrightarrow V_{\epsilon_j}$ to be the identity map.
\begin{proposition}
Let $a(\epsilon)=\frac{A}{\epsilon}, b(\epsilon)=-\frac{A}{\epsilon}$. Then   $\eta^{\mathfrak{h}}_A$ is  the strong contraction of the family of representations $\left\{        ( \rho^{\mathfrak{ea}}_{(a(\epsilon),b(\epsilon),\epsilon)},V_{\epsilon}   )      \right\}_{\epsilon \in I_1}$ with respect to the contraction $\mathfrak{ea}\stackrel{t(\epsilon)}{\rightarrow} \mathfrak{h}$.
\label{prop1}
\end{proposition}

\textit{proof}.
Let $V$ be $L_c^{2,\infty}(\mathbb{R},dx)$. Since the  embeddings $\varphi_{\epsilon_i\epsilon_j}$ are the identity map they  are obviously compatible and satisfy conditions $(i, ii)$ and  $(iii)$ in definition \ref{def3} and also $L(f)=f$ for every $f(x) \in L_c^{2,\infty}(\mathbb{R},dx)$. For condition $(iv)$ we observe that for  $f \in V_{\epsilon_1}=L_c^{2,\infty}(\mathbb{R},dx)$ we have:
\begin{eqnarray}
&&\lim_{\epsilon \longrightarrow 0^+}  L(\rho^{\mathfrak{ea}}_{(a(\epsilon),b(\epsilon),\epsilon)}(t_{\epsilon}(X_{3}^{\mathfrak{ea}}))\varphi_{\epsilon_1\epsilon}f)(x)= \\ \nonumber
&&\lim_{\epsilon \longrightarrow 0^+}  \rho^{\mathfrak{ea}}_{(a(\epsilon),b(\epsilon),\epsilon)}(-\epsilon X_3^{\mathfrak{ea}})f(x)    = \lim_{\epsilon \longrightarrow 0} \epsilon \frac{1}{\epsilon}\frac{d}{dx}f(x)=\frac{d}{dx}f(x)=\\ \nonumber
&& \frac{d}{dx}L(f)(x)=\eta^{\mathfrak{h}}_{A}(\psi(X_{3}^{\mathfrak{ea}}))L(f)(x)
\end{eqnarray}
\begin{eqnarray}
&& \lim_{\epsilon \longrightarrow 0^+}  L(\rho^{\mathfrak{ea}}_{(a(\epsilon),b(\epsilon),\epsilon)}(t_{\epsilon}(X_1^{\mathfrak{ea}}+X_2^{\mathfrak{ea}}))\varphi_{\epsilon_1\epsilon}f)(x)= \\ \nonumber
&& \lim_{\epsilon \longrightarrow 0^+}  \rho^{\mathfrak{ea}}_{(a(\epsilon),b(\epsilon),\epsilon)}(X_1^{\mathfrak{ea}}+X_2^{\mathfrak{ea}})f(x)=\lim_{\epsilon \longrightarrow 0^+} i(b(\epsilon)e^{-\epsilon x}+a(\epsilon))f(x) =\\ \nonumber
&& \lim_{\epsilon \longrightarrow 0^+} i((-\frac{A}{\epsilon})(1-\epsilon x)+\frac{A}{\epsilon})f(x)=iAxf(x)=  iAxL(f)(x)=  \\ \nonumber
&& \eta^{\mathfrak{h}}_{A}(\psi(X_1^{\mathfrak{ea}}+X_2^{\mathfrak{ea}}))L(f)(x)
\end{eqnarray}
\begin{eqnarray}
&& \lim_{\epsilon \longrightarrow 0^+}  L(\rho^{\mathfrak{ea}}_{(a(\epsilon),b(\epsilon),\epsilon)}(t_{\epsilon}(X_{2}^{\mathfrak{ea}}))\varphi_{\epsilon_1\epsilon}f)(x)=\\ \nonumber
&&\lim_{\epsilon \longrightarrow 0^+}   \rho^{\mathfrak{ea}}_{(a(\epsilon),b(\epsilon),\epsilon)}(\epsilon X_2^{\mathfrak{ea}})f(x) = \lim_{\epsilon \longrightarrow 0^+} i\epsilon a(\epsilon))f(x)= \lim_{\epsilon \longrightarrow 0^+} i\epsilon (\frac{A}{\epsilon})f(x)= \\ \nonumber
&& iAf(x)=iAL(f)(x)=\eta^{\mathfrak{h}}_{A}(\psi(X_{2}^{\mathfrak{ea}}))L(f)(x)
\end{eqnarray}
The above pointwise limit can be shown to be a uniform limit due to the compactness of the support of $f$ and this implies that we also have the desired convergence in norm.
\subsection{The contraction: $iso(2)\longrightarrow \mathfrak{h}$}
\label{C2}Set $\g=iso(2)$. The maps
\begin{eqnarray}
&& t_{\epsilon}:iso(2)\longrightarrow iso(2) \\ \nonumber
&& t_{\epsilon}(X_2^{iso(2)})=\epsilon X_2^{iso(2)} \\ \nonumber
&& t_{\epsilon}(X_1^{iso(2)})=X_1^{iso(2)}\\ \nonumber
&& t_{\epsilon}(X_3^{iso(2)})=\epsilon X_3^{iso(2)}
\end{eqnarray}
realize the contraction of the Lie algebra and lead to $[X_3,X_1]_0=X_2$
which induce the isomorphism
\begin{eqnarray}
&& \psi:\g_0\longrightarrow \mathfrak{h} \\ \nonumber
&& \psi(X_2^{iso(2)})= X_1^{\mathfrak{h}} \\ \nonumber
&& \psi(X_1^{iso(2)})=X_2^{\mathfrak{h}}  \\ \nonumber
&& \psi(X_3^{iso(2)})= X_3^{\mathfrak{h}}
\end{eqnarray}
Using the Mackey machine one can show that for every $(r_1,r_2) \in \mathbb{R}^2$ such that $r_1^2+r_2^2\neq0$ the map
\begin{eqnarray}
&& \rho^{iso(2)}_{(r_1,r_2)}:iso(2) \longrightarrow gl(L^{2,\infty}_{0}([-\pi,\pi],dx)) \\ \nonumber
&& \rho^{iso(2)}_{(r_1,r_2)}(X_1^{iso(2)})= i(r_1\sin{x}+r_2\cos{x})\\ \nonumber
&&   \rho^{iso(2)}_{(r_1,r_2)}(X_2^{iso(2)})= i(r_1\cos{x}-r_2\sin{x}) \\ \nonumber
&& \rho^{iso(2)}_{(r_1,r_2)}(X_3^{iso(2)})= \frac{d}{dx}  \nonumber
\end{eqnarray}
is a SHIIR, where $L^{2,\infty}_{0}([-\pi,\pi],dx)$ stands for the inner product space of smooth functions on $[-\pi,\pi]$ with the standard inner product and such that for all $k\in \mathbb{N}_0$  $f^{(k)}(\pm \pi)=0$. In fact, these are all the SHIIR of $iso(2)$ except for the one dimensional ones.
For every $\epsilon \in I_1$ let $V_{\epsilon}$ be the inner product space $L^{2,\infty}_0([-\frac{ \pi}{ \epsilon},\frac{\pi}{\epsilon}],dx)$ and we define the function $\psi_{\epsilon}:[-\frac{ \pi}{ \epsilon},\frac{\pi}{\epsilon}]\longrightarrow [-\pi,\pi]$ by $\psi_{\epsilon}(x)=\epsilon x$.
We note that $P_{\epsilon}:L^{2,\infty}_0([-\pi,\pi],dx)\longrightarrow V_{\epsilon}$ which is given by $P_{\epsilon}(f)=f \circ \psi_{\epsilon}$, is  an isomorphism. Its inverse $P_{\epsilon}^{-1}$ is given by  $P_{\epsilon}^{-1}(f)=f \circ \psi^{-1}_{\epsilon}$.
For $\epsilon_i, \epsilon_j\in I_1$ such that $\epsilon_i \geq \epsilon_j$ we define the linear transformation   $\varphi_{\epsilon_i,\epsilon_j}:V_{\epsilon_i}\longrightarrow V_{\epsilon_j}$ to be the inclusion map that is defined as follows: for any $f\in V_{\epsilon_i}$, $\varphi_{\epsilon_i,\epsilon_j}(f)(x)=f(x)$ for  $x \in [-\frac{ \pi}{ \epsilon_i},\frac{\pi}{\epsilon_i}]$ and $\varphi_{\epsilon_i,\epsilon_j}(f)(x)=0$ for $x \in [-\frac{ \pi}{ \epsilon_j},\frac{\pi}{\epsilon_j}]  \backslash [-\frac{ \pi}{ \epsilon_i},\frac{\pi}{\epsilon_i}]$.
This function obviously preserves the inner product and satisfies the compatibility condition.
For each  $\epsilon \in I_1$   we intertwine  $\rho^{iso(2)}_{(r_1,r_2)}$ with $P_{\epsilon}$ to get the equivalent representation  $\rho^{iso(2)}_{(r_1,r_2,\epsilon)}$ given by:
\begin{eqnarray}
&& \rho^{iso(2)}_{(r_1,r_2,\epsilon)}:iso(2)\longrightarrow gl(V_{\epsilon}) \\ \nonumber
&&  \rho^{iso(2)}_{(r_1,r_2,\epsilon)}(X_1^{iso(2)})  =  P_{\epsilon} \circ \rho^{iso(2)}_{(r_1,r_2)}(X_1^{iso(2)}) \circ P^{-1}_{\epsilon} =i(r_1\sin{\epsilon x}+r_2\cos{\epsilon x})  \\ \nonumber
&& \rho^{iso(2)}_{(r_1,r_2,\epsilon)}(X_2^{iso(2)})= P_{\epsilon} \circ \rho^{iso(2)}_{(r_1,r_2)}(X_2^{iso(2)}) \circ P^{-1}_{\epsilon}=i(r_1\cos{\epsilon x}-r_2\sin{\epsilon x}) \\ \nonumber
&& \rho^{iso(2)}_{(r_1,r_2,\epsilon)}(X_3^{iso(2)})=P_{\epsilon} \circ \rho^{iso(2)}_{(r_1,r_2)}(X_3^{iso(2)}) \circ P^{-1}_{\epsilon}=\frac{1}{\epsilon}\frac{d}{dx}  \nonumber
\end{eqnarray}
\begin{proposition}
Let $r_1(\epsilon)=\frac{A}{\epsilon}, r_2(\epsilon)=0$. Then   $\eta^{\mathfrak{h}}_A$ is  the strong contraction of the family of representations $\left\{        ( \rho^{iso(2)}_{(r_1(\epsilon),r_2(\epsilon),\epsilon)},V_{\epsilon}   )      \right\}_{\epsilon \in I_1}$ with respect to the contraction $iso(2)\stackrel{t(\epsilon)}{\rightarrow} \mathfrak{h}$.
\label{prop2}
\end{proposition}
\textit{proof}.
Let  $V$ be $L^{2,\infty}_c(\mathbb{R},dx)$. 	For every $f\in V_{\epsilon}$, $L(f)(x)=f(x)$ for $x \in [-\frac{ \pi}{ \epsilon},\frac{\pi}{\epsilon}]$ and zero otherwise.  
It is easy to see that$(i), (ii)$ and $(ii)$ of definition \ref{def3} are satisfied. For $(iv)$ we observe that for $f\in V_{\epsilon_0}$ and $x$ in the support of $f$  we have:
\begin{eqnarray}
&&\lim_{\epsilon \longrightarrow 0^+}  L(\rho^{iso(2)}_{(r_1(\epsilon),r_2(\epsilon),\epsilon)}(t_{\epsilon}(X_{3}^{iso(2)}))\varphi_{\epsilon_0,\epsilon}(f))(x)= \\ \nonumber
&& \lim_{\epsilon \longrightarrow 0^+}  \epsilon \frac{1}{\epsilon}\frac{d}{dx}(f)(x) = f'(x)=L(f')(x)=\frac{d}{dx}(L(f))(x)=\\ \nonumber
&& \eta_A^{\mathfrak{h}}(\psi(X_{3}^{iso(2)})) L(f)(x)
\end{eqnarray}
\begin{eqnarray}
&& \lim_{\epsilon \longrightarrow 0^+}  L(\rho^{iso(2)}_{(r_1(\epsilon),r_2(\epsilon),\epsilon)}(t_{\epsilon}(X_{1}^{iso(2)}))\varphi_{\epsilon_0,\epsilon}(f))(x)=  \\ \nonumber
&&\lim_{\epsilon \longrightarrow 0^+} i\frac{A}{\epsilon}\sin(\epsilon x)f(x)=iAxf(x)=iAxL(f)(x) =  \eta_A^{\mathfrak{h}}(\psi(X_{1}^{iso(2)}))L(f)(x)
\end{eqnarray}
\begin{eqnarray}
&& \lim_{\epsilon \longrightarrow 0^+}  L(\rho^{iso(2)}_{(r_1(\epsilon),r_2(\epsilon),\epsilon)}(t_{\epsilon}(X_{2}^{iso(2)}))\varphi_{\epsilon_0,\epsilon}(f))(x)=  \\ \nonumber
&&\lim_{\epsilon \longrightarrow 0^+}  i\epsilon(\frac{A}{\epsilon}\cos(\epsilon x) (f))(x) =iAf(x)=iAL(f)= \eta_A^{\mathfrak{h}}(\psi(X_{2}^{iso(2)}))L(f)(x)
\end{eqnarray}

The above pointwise limit can be shown to be a uniform limit due to the compactness of the support of $f$ and this implies that we also have the desired convergence in norm.\newline
\textbf{Remark 2.}
\textit{Since the support of any Hermite function is not compact the canonical basis for the representations of the Heisenberg Lie algebra which consists of Hermite functions does not  belong to our limit space.  It seems  that one can not find a compatible (see appendix) family of bases such that the matrix elements of these bases converge under contraction to those of the canonical basis. }

\subsection{The contraction: $\mathfrak{g}(\lambda)_{\lambda \neq 1}\longrightarrow \mathfrak{h}$}
Set $\g=\mathfrak{g}(\lambda)_{\lambda \neq 1}$. The maps
\begin{eqnarray}
&& t_{\epsilon}:\mathfrak{g}(\lambda)\longrightarrow \mathfrak{g}(\lambda) \\ \nonumber
&& t_{\epsilon}(X_1^{\mathfrak{g}(\lambda)})=\epsilon (1-\lambda)X_1^{\mathfrak{g}(\lambda)} \\ \nonumber
&& t_{\epsilon}(X_1^{\mathfrak{g}(\lambda)}+X_2^{\mathfrak{g}(\lambda)})=X_1^{\mathfrak{g}(\lambda)}+X_2^{\mathfrak{g}(\lambda)}\\ \nonumber
&& t_{\epsilon}(X_3^{\mathfrak{g}(\lambda)})=\epsilon X_3^{\mathfrak{g}(\lambda)}
\end{eqnarray}
realize the contraction of the Lie algebra  and lead to $[X_3,X_1+X_2]_0=X_1$
which induce the isomorphism
\begin{eqnarray}
&& \psi:\g_0\longrightarrow \mathfrak{h} \\ \nonumber
&& \psi(X_1^{\mathfrak{g}(\lambda)})= X_1^{\mathfrak{h}} \\ \nonumber
&& \psi(X_1^{\mathfrak{g}(\lambda)}+X_2^{\mathfrak{g}(\lambda)})=X_2^{\mathfrak{h}}  \\ \nonumber
&& \psi(X_3^{\mathfrak{g}(\lambda)})= X_3^{\mathfrak{h}}
\end{eqnarray}
Using the Mackey machine one can show that for every $a,b\in \mathbb{R}$ such that $a^2+b^2\neq 0$ the map
\begin{eqnarray}
&& \rho^{\mathfrak{g}(\lambda)}_{(a,b)}:{\mathfrak{g}(\lambda)} \longrightarrow gl(L^{2,\infty}_c(\mathbb{R},dx)) \\ \nonumber
&& \rho^{\mathfrak{g}(\lambda)}_{(a,b)}(X_1^{\mathfrak{g}(\lambda)})= iae^x \\ \nonumber
&& \rho^{\mathfrak{g}(\lambda)}_{(a,b)}(X_2^{\mathfrak{g}(\lambda)})= ibe^{\lambda x} \\ \nonumber
&& \rho^{\mathfrak{g}(\lambda)}_{(a,b)}(X_3^{\mathfrak{g}(\lambda)})= \frac{d}{dx}  \nonumber
\end{eqnarray}
is a SHIIR. In fact for every real value of $\lambda$ these are all the SHIIR of $\mathfrak{g}(\lambda)$ except for the one dimensional ones.
For every $\epsilon \in I_1$ let $V_{\epsilon}$ be the inner product space $L_c^{2,\infty}(\mathbb{R},dx)$ and we define a function $\psi_{\epsilon}:\mathbb{R}\longrightarrow \mathbb{R}$ by $\psi_{\epsilon}(x)=\epsilon x$.
We note that $P_{\epsilon}:L_c^{2,\infty}(\mathbb{R},dx)\longrightarrow V_{\epsilon}$ which is given by $P_{\epsilon}(f)=f \circ \psi_{\epsilon}$ is  an isomorphism. Its inverse $P_{\epsilon}^{-1}$ is given by  $P_{\epsilon}^{-1}(f)=f \circ \psi^{-1}_{\epsilon}$.
For each  $\epsilon \in I_1$  and $(a,b)\in \mathbb{R} \times \mathbb{R}^{*}$ we intertwine  $\rho^{\mathfrak{g}(\lambda)}_{(a,b)}$ with $P_{\epsilon}$ to get the equivalent  representation $\rho^{\mathfrak{g}(\lambda)}_{(a,b,\epsilon)}$ given by:
\begin{eqnarray}
&& \rho^{\mathfrak{g}(\lambda)}_{(a,b,\epsilon)}:\mathfrak{g}(\lambda)\longrightarrow gl(V_{\epsilon})\\ \nonumber
&&  \rho^{\mathfrak{g}(\lambda)}_{(a,b,\epsilon)}(X_1^{\mathfrak{g}(\lambda)})  =  P_{\epsilon} \circ \rho^{\mathfrak{g}(\lambda)}_{(a,b)}(X_1^{\mathfrak{g}(\lambda)}) \circ P^{-1}_{\epsilon} =iae^{\epsilon x}  \\ \nonumber
&& \rho^{\mathfrak{g}(\lambda)}_{(a,b,\epsilon)}(X_2^{\mathfrak{g}(\lambda)})= P_{\epsilon} \circ \rho^{\mathfrak{g}(\lambda)}_{(a,b)}(X_2^{\mathfrak{g}(\lambda)}) \circ P^{-1}_{\epsilon}=ibe^{\lambda \epsilon x} \\ \nonumber
&& \rho^{\mathfrak{g}(\lambda)}_{(a,b,\epsilon)}(X_3^{\mathfrak{g}(\lambda)})=P_{\epsilon} \circ \rho^{\mathfrak{g}(\lambda)}_{(a,b)}(X_3^{\mathfrak{g}(\lambda)}) \circ P^{-1}_{\epsilon}=\frac{1}{\epsilon}\frac{d}{dx}  \nonumber
\end{eqnarray}
For $\epsilon_i, \epsilon_j\in I_1$ such that $\epsilon_j \leq \epsilon_i$ we define the linear isometry  $\varphi_{\epsilon_i,\epsilon_j}:V_{\epsilon_i}\longrightarrow V_{\epsilon_j}$ to be the identity map. These maps are obviously compatible.

\begin{proposition}
Let $a(\epsilon)=\frac{A}{\epsilon(1-\lambda)}, b(\epsilon)=-\frac{A}{\epsilon(1-\lambda)}$. Then   $\eta^{\mathfrak{h}}_A$ is  the strong contraction of the family of representations $\left\{        ( \rho^{\mathfrak{g}(\lambda)}_{(a(\epsilon),b(\epsilon),\epsilon)},V_{\epsilon}   )      \right\}_{\epsilon \in I_1}$ with respect to the contraction $\mathfrak{g}(\lambda)_{\lambda \neq 1}\stackrel{t(\epsilon)}{\rightarrow} \mathfrak{h}$.
\end{proposition}
The proof is similar to that of proposition \ref{prop1}.

\subsection{The contraction: $\mathfrak{l}(\lambda)_{\lambda\neq 0}\longrightarrow \mathfrak{h}$}
Set $\g=\mathfrak{l}(\lambda)_{\lambda\neq 0}$. The maps
\begin{eqnarray}
&& t_{\epsilon}:\mathfrak{l}(\lambda)\longrightarrow \mathfrak{l}(\lambda) \\ \nonumber
&& t_{\epsilon}(X_2^{\mathfrak{l}(\lambda)})=\epsilon X_2^{\mathfrak{l}(\lambda)} \\ \nonumber
&& t_{\epsilon}(X_1^{\mathfrak{l}(\lambda)})=X_1^{\mathfrak{l}(\lambda)}\\ \nonumber
&& t_{\epsilon}(X_3^{\mathfrak{l}(\lambda)} )=\epsilon X_3^{\mathfrak{l}(\lambda)}
\end{eqnarray}
realize the contraction of the Lie algebra and lead to $[X_3,X_1]_0=X_2$
which induce the isomorphism
\begin{eqnarray}
&& \psi:\g_0\longrightarrow \mathfrak{h} \\ \nonumber
&& \psi(X_1^{\mathfrak{g}(\lambda)})= X_1^{\mathfrak{h}} \\ \nonumber
&& \psi(X_1^{\mathfrak{g}(\lambda)}+X_2^{\mathfrak{g}(\lambda)})=X_2^{\mathfrak{h}}  \\ \nonumber
&& \psi(X_3^{\mathfrak{g}(\lambda)})= X_3^{\mathfrak{h}}
\end{eqnarray}
Using the Mackey machine one can show that for every $a,b\in \mathbb{R}$ such that $a^2+b^2\neq 0$ the map
\begin{eqnarray}
&& \rho^{\mathfrak{l}(\lambda)}_{(a,b)}:{\mathfrak{l}(\lambda)} \longrightarrow gl(L^{2,\infty}_c(\mathbb{R},dx)) \\ \nonumber
&& \rho^{\mathfrak{l}(\lambda)}_{(a,b)}(X_1^{\mathfrak{l}(\lambda)})= iae^{\lambda x}\cos{x}+ibe^{\lambda x}\sin{x} \\ \nonumber
&&   \rho^{\mathfrak{l}(\lambda)}_{(a,b)}(X_2^{\mathfrak{l}(\lambda)})= -iae^{\lambda x}\sin{x}+ibe^{\lambda x}\cos{x} \\ \nonumber
&& \rho^{\mathfrak{l}(\lambda)}_{(a,b)}(X_3^{\mathfrak{l}(\lambda)})= \frac{d}{dx}  \nonumber
\end{eqnarray}
is a SHIIR.
Using the same intertwiner  as in  the contraction  $\mathfrak{g}(\lambda)_{\lambda \neq 1}\longrightarrow \mathfrak{h}$ we obtain the equivalent representation $\rho^{\mathfrak{l}(\lambda)}_{(a,b,\epsilon)}$ given by:

\begin{eqnarray}
&&  \rho^{\mathfrak{l}(\lambda)}_{(a,b,\epsilon)}(X_1^{\mathfrak{l}(\lambda)})  =  iae^{\lambda \epsilon x}\cos{\epsilon x}+ibe^{\lambda \epsilon x}\sin{\epsilon x}  \\ \nonumber
&& \rho^{\mathfrak{l}(\lambda)}_{(a,b,\epsilon)}(X_2^{\mathfrak{l}(\lambda)})=-iae^{\lambda \epsilon x}\sin{\epsilon x}+ibe^{\lambda \epsilon x}\cos{\epsilon x} \\ \nonumber
&& \rho^{\mathfrak{l}(\lambda)}_{(a,b,\epsilon)}(X_3^{\mathfrak{l}(\lambda)})=\frac{1}{\epsilon}\frac{d}{dx}  \nonumber
\end{eqnarray}
We define   isometric embeddings  as in the case of $\mathfrak{g}(\lambda)_{\lambda \neq 1}\longrightarrow \mathfrak{h}$.
\begin{proposition}
Let $a(\epsilon)=0, b(\epsilon)=\frac{A}{\epsilon}$. Then   $\eta^{\mathfrak{h}}_A$ is  the strong contraction of the family of representations $\left\{        ( \rho^{\mathfrak{l}(\lambda)}_{(a(\epsilon),b(\epsilon),\epsilon)},V_{\epsilon}   )      \right\}_{\epsilon \in I_1}$ with respect to the contraction $\mathfrak{l}(\lambda)_{\lambda\neq 0}\stackrel{t(\epsilon)}{\rightarrow} \mathfrak{h}$.
\end{proposition}
The proof is similar to that of proposition \ref{prop1}.

\subsection{The contraction: $\mathfrak{c}\longrightarrow \mathfrak{h}$}
Set $\g=\mathfrak{c}$. The maps
\begin{eqnarray}
&& t_{\epsilon}:\mathfrak{c}\longrightarrow \mathfrak{c} \\ \nonumber
&& t_{\epsilon}(X_1^{\mathfrak{c}})=\epsilon X_1^{\mathfrak{c}} \\ \nonumber
&& t_{\epsilon}(X_2^{\mathfrak{c}})=X_2^{\mathfrak{c}}\\ \nonumber
&& t_{\epsilon}(X_3^{\mathfrak{c}})=\epsilon X_3^{\mathfrak{c}}
\end{eqnarray}
realize the contraction of the Lie algebra  and lead to $[X_3,X_2]_0=X_1$
which induce the isomorphism
\begin{eqnarray}
&& \psi:\g_0\longrightarrow \mathfrak{h} \\ \nonumber
&& \psi(X_1^{\mathfrak{c}})= X_1^{\mathfrak{h}} \\ \nonumber
&& \psi(X_2^{\mathfrak{c}})=X_2^{\mathfrak{h}}  \\ \nonumber
&& \psi(X_3^{\mathfrak{c}})= X_3^{\mathfrak{h}}
\end{eqnarray}
Using the  Mackey machine one can show that for every $a,b\in \mathbb{R}$ such that $a^2+b^2\neq 0$ the map
\begin{eqnarray}
&& \rho^{\mathfrak{c}}_{(a,b)}:\mathfrak{c} \longrightarrow gl(L^{2,\infty}_c(\mathbb{R},dx)) \\ \nonumber
&& \rho^{\mathfrak{c}}_{(a,b)}(X_1^{\mathfrak{c}})= iae^{ x} \\ \nonumber
&&   \rho^{\mathfrak{c}}_{(a,b)}(X_2^{\mathfrak{c}})= iae^{ x}x+ibe^{ x} \\ \nonumber
&& \rho^{\mathfrak{c}}_{(a,b)}(X_3^{\mathfrak{c}})= \frac{d}{dx}  \nonumber
\end{eqnarray}
is a SHIIR and in fact these are all its SHIIR except for the one dimensional ones.\newline
Using the same intertwiner as in the  contraction  $\mathfrak{g}(\lambda)_{\lambda \neq 1}\longrightarrow \mathfrak{h}$ we obtain the equivalent representation $\rho^{\mathfrak{c}}_{(a,b,\epsilon)}$ given by:
\begin{eqnarray}
&& \rho^{\mathfrak{c}}_{(a,b,\epsilon)}:\mathfrak{c}\longrightarrow gl(V_{\epsilon})\\ \nonumber
&&  \rho^{\mathfrak{c}}_{(a,b,\epsilon)}(X_1^{\mathfrak{c}})  =  P_{\epsilon} \circ \rho^{\mathfrak{c}}_{(a,b)}(X_1^{\mathfrak{c}}) \circ P^{-1}_{\epsilon} =iae^{\epsilon x}  \\ \nonumber
&& \rho^{\mathfrak{c}}_{(a,b,\epsilon)}(X_2^{\mathfrak{c}})= P_{\epsilon} \circ \rho^{\mathfrak{c}}_{(a,b)}(X_2^{\mathfrak{c}}) \circ P^{-1}_{\epsilon}=iae^{\epsilon x}\epsilon x+ibe^{ \epsilon x} \\ \nonumber
&& \rho^{\mathfrak{c}}_{(a,b,\epsilon)}(X_3^{\mathfrak{c}})=P_{\epsilon} \circ \rho^{\mathfrak{c}}_{(a,b)}(X_3^{\mathfrak{c}}) \circ P^{-1}_{\epsilon}=\frac{1}{\epsilon}\frac{d}{dx}  \nonumber
\end{eqnarray}
We define isometric embeddings  as in the case of $\mathfrak{g}(\lambda)_{\lambda \neq 1}\longrightarrow \mathfrak{h}$.
\begin{proposition}
Let $a(\epsilon)=\frac{A}{\epsilon}, b(\epsilon)=0$. Then   $\eta^{\mathfrak{h}}_A$ is  the strong contraction of the family of representations $\left\{        ( \rho^{\mathfrak{c}}_{(a(\epsilon),b(\epsilon),\epsilon)},V_{\epsilon}   )      \right\}_{\epsilon \in I_1}$ with respect to the contraction $\mathfrak{c}\stackrel{t(\epsilon)}{\rightarrow} \mathfrak{h}$.
\end{proposition}
The proof is similar to that of proposition \ref{prop1}.

\subsection{The contraction: $\mathfrak{c}\longrightarrow \mathfrak{g}(1)$}
Set $\g=\mathfrak{c}$. The maps
\begin{eqnarray}
&& t_{\epsilon}:\mathfrak{c}\longrightarrow \mathfrak{c} \\ \nonumber
&& t_{\epsilon}(X_1^{\mathfrak{c}})= X_1^{\mathfrak{c}} \\ \nonumber
&& t_{\epsilon}(X_2^{\mathfrak{c}})=\epsilon X_2^{\mathfrak{c}}\\ \nonumber
&& t_{\epsilon}(X_3^{\mathfrak{c}})= X_3^{\mathfrak{c}}
\end{eqnarray}
realize the contraction of the Lie algebra and lead to $[X_3,X_1]_0=X_1, [X_3,X_2]_0= X_2$
which induce the isomorphism
\begin{eqnarray}
&& \psi:\g_0\longrightarrow \mathfrak{g}(1) \\ \nonumber
&& \psi(X_1^{\mathfrak{c}})= X_1^{\mathfrak{g}(1)} \\ \nonumber
&& \psi(X_2^{\mathfrak{c}})=X_2^{\mathfrak{g}(1)}  \\ \nonumber
&& \psi(X_3^{\mathfrak{c}})= X_3^{\mathfrak{g}(1)}
\end{eqnarray}
Let $V_{\epsilon}$ and $ \varphi_{\epsilon_i,\epsilon_j}$ be as in the contraction $\mathfrak{g}(\lambda)_{\lambda \neq 1}\longrightarrow \mathfrak{h}$.
\begin{proposition}
Let $a(\epsilon)=a, b(\epsilon)=\frac{b}{\epsilon}$. Then   $\rho^{\mathfrak{g}(1)}_{(a,b)}$ is  the strong contraction of the family of representations $\left\{        ( \rho^{\mathfrak{c}}_{(a(\epsilon),b(\epsilon))},V_{\epsilon}   )      \right\}_{\epsilon \in I_1}$ with respect to the contraction $\mathfrak{c}\stackrel{t(\epsilon)}{\rightarrow} \mathfrak{g}(1)$.
\end{proposition}
The proof is similar to that of proposition \ref{prop1}.

\subsection{The contraction: $su(2)\longrightarrow iso(2)$}
\label{C7}
Set $\g=su(2)$. The maps
\begin{eqnarray}
&& t_{\epsilon}:su(2)\longrightarrow su(2) \\ \nonumber
&& t_{\epsilon}(X_1^{su(2)})=\epsilon X_1^{su(2)} \\ \nonumber
&& t_{\epsilon}(X_2^{su(2)})=\epsilon X_2^{su(2)}\\ \nonumber
&& t_{\epsilon}(X_3^{su(2)})= X_3^{su(2)}
\end{eqnarray}
realize the contraction of the Lie algebra and lead to $[X_3,X_1]_0=X_2, [X_3,X_2]_0=-X_1$
which induce the isomorphism
\begin{eqnarray}
&& \psi:\g_0\longrightarrow iso(2) \\ \nonumber
&& \psi(X_1^{su(2)})= -X_2^{iso(2)} \\ \nonumber
&& \psi(X_2^{su(2)})=X_1^{iso(2)}  \\ \nonumber
&& \psi(X_3^{su(2)})= X_3^{iso(2)}
\end{eqnarray}
For every $l \in \mathbb{N}_0$ let $\mathbb{H}_l$ be the space of spherical harmonics of degree $l$ with the inner product
$$\left\langle f,g\right\rangle=\int_{0}^{2\pi}\int_{0}^{\pi}f(\theta,\varphi)\overline{g(\theta,\varphi)}(2l+1)\sin(\theta)d\theta d\varphi$$
For any such $l$ the map
\begin{eqnarray}
&& \rho^{su(2)}_{l}:su(2) \longrightarrow gl(\mathbb{H}_l) \\ \nonumber
&& \rho^{su(2)}_{l}(X_1^{su(2)})= \sin{\varphi}\frac{d}{d\theta}+\cot{\theta}\cos{\varphi}\frac{d}{d\varphi}\\ \nonumber
&& \rho^{su(2)}_{l}(X_2^{su(2)})= -\cos{\varphi}\frac{d}{d\theta}+\cot{\theta}\sin{\varphi}\frac{d}{d\varphi} \\ \nonumber
&& \rho^{su(2)}_{l}(X_3^{su(2)})= -\frac{d}{d\varphi}  \nonumber
\end{eqnarray}
is a SHIIR \cite{Vil}.
For every $\epsilon\in (0,\infty)$ we define the $\epsilon$-deformed disc to be $S^2_{\epsilon}= [0,\frac{\pi}{\epsilon}) \times [0,2\pi) $. We also define a bijection $\psi_{\epsilon}:S^2_{\epsilon}\longrightarrow S^2_{1}$ by $\psi_{\epsilon}(\theta,\varphi)=(\epsilon\theta,\varphi)$.
For every $l\in \mathbb{N}$ and every $\epsilon\in (0,\infty)$ we define $\mathbb{H}_{(l,\epsilon)}$  to be the space of spherical harmonics on $S^2_{\epsilon}$ which is defined to be the image of $\mathbb{H}_l$ under the map $P_{\epsilon}$ that takes $f\in \mathbb{H}_l$ to $f\circ \psi_{\epsilon}$ and we endowed $\mathbb{H}_{(l,\epsilon)}$ with an inner product which is defined by
$$\left\langle f,g\right\rangle_{(l,\epsilon)}=\int_{0}^{2\pi}\int_{0}^{\frac{\pi}{\epsilon}}f(\theta,\varphi)\overline{g(\theta,\varphi)}(2l+1)\sin(\epsilon \theta)\epsilon d\theta d\varphi$$
One can easily check that $P_{\epsilon}$ is an isomorphism of vector spaces. The functions $\chi_{l,\epsilon}^m(\theta,\varphi)\equiv Y_{l}^m(\epsilon \theta,\varphi)$ where $m$ is ranging over $-l, -l+1,...l$ and $Y_{l}^m(\theta,\varphi)$ are the spherical harmonics (we use the convention of \cite{Vil}) form a basis for $\mathbb{H}_{(l,\epsilon)}$.
For each  $\epsilon\in (0,\infty)$  we intertwine  $\rho^{su(2)}_{l}$ with $P_{\epsilon}$ to get the equivalent representation  $\rho^{su(2)}_{(l,\epsilon)}$ given by:
\begin{eqnarray}
&&\rho^{su(2)}_{(l,\epsilon)}:su(2)\longrightarrow gl(\mathbb{H}_{(l,\epsilon)}) \\ \nonumber
&&  \rho^{su(2)}_{(l,\epsilon)}(X_1^{su(2)})=\sin{\varphi}\frac{1}{\epsilon}\frac{d}{d\theta}+\cot{\epsilon \theta}\cos{\varphi}\frac{d}{d\varphi}\\ \nonumber
&&  \rho^{su(2)}_{(l,\epsilon)}(X_2^{su(2)})=-\cos{\varphi}\frac{1}{\epsilon}\frac{d}{d\theta}+\cot{\epsilon \theta}\sin{\varphi}\frac{d}{d\varphi} \\ \nonumber
&&  \rho^{su(2)}_{(l,\epsilon)}(X_3^{su(2)})= -\frac{d}{d\varphi}  \nonumber
\end{eqnarray}
The action of the generators on the basis element $\chi_{l,\epsilon}^m(\theta,\varphi,)$ is given by:
\begin{eqnarray}\label{eq41}
&   \rho^{su(2)}_{(l,\epsilon)}(X_1^{su(2)})\chi_{l,\epsilon}^m  =& \frac{1}{2}(\sqrt{(l+m)(l-m+1)}\chi_{l,\epsilon}^{m-1}-\\ \nonumber
&&\sqrt{(l-m)(l+m+1)}\chi_{l,\epsilon}^{m+1}) \\ \nonumber
&  \rho^{su(2)}_{(l,\epsilon)}(X_2^{su(2)})\chi_{l,\epsilon}^m  =& \frac{-i}{2}(\sqrt{(l+m)(l-m+1)}\chi_{l,\epsilon}^{m-1}+\\ \nonumber
&&\sqrt{(l-m)(l+m+1)}\chi_{l,\epsilon}^{m+1})\\ \nonumber
&  \rho^{su(2)}_{(l,\epsilon)}(X_3^{su(2)})\chi_{l,\epsilon}^m  =& im\chi_{l,\epsilon}^m  \nonumber
\end{eqnarray}
For every $0 < R\in \mathbb{R}$ we define $\epsilon^R_l=\frac{R}{l}$. Let $V_l$ be the inner product space $\mathbb{H}_{(l,\epsilon^R_l)}$ and for $i,j\in \mathbb{N}_0$ such that $i \leq j$ we define  $\varphi_{i,j}:V_{i}\longrightarrow V_{j}$ to be the linear map that satisfies $\varphi_{i,j}(\chi_{i,\epsilon_i^R}^m)=\chi_{j,\epsilon_j^R}^m$. $\varphi_{i,j}$ are compatible isometric injections.
We now give a realization of the SHIIR of $iso(2)$ which is taken from \cite{Vil} and is more suitable for the contraction $su(2)\stackrel{t(\epsilon)}{\longrightarrow}iso(2)$. For every $0 \neq R\in \mathbb{R}$ let $\Omega_R$ be the the subspace of $L^{2,\infty}(\mathbb{R}^2,dx)$ which has the basis $\left\{B^R_m(r,\varphi)=(i)^mJ_m(Rr)e^{im\varphi}|m\in \mathbb{Z} \right\}$. On $\Omega_R$  the SHIIR of $iso(2)$ in polar coordinates is given by:
\begin{eqnarray}
&& \eta^{iso(2)}_{R}:iso(2)\longrightarrow gl(\Omega_R) \\ \nonumber
&&  \eta^{iso(2)}_{R}(X_1^{iso(2)})  = -\cos{\varphi}\frac{d}{dr}+\frac{\sin{\varphi}}{r}\frac{d}{d\varphi}  \\ \nonumber
&&  \eta^{iso(2)}_{R}(X_2^{iso(2)})  = -\sin{\varphi}\frac{d}{dr}-\frac{\cos{\varphi}}{r}\frac{d}{d\varphi}\\ \nonumber
&&  \eta^{iso(2)}_{r}(X_3^{iso(2)})  =   -\frac{d}{d\varphi}  \nonumber
\end{eqnarray}
We remark that the representation $\eta^{iso(2)}_{R}$ is equivalent to every one of the representations $\rho^{\mathfrak{l}(0)}_{(a,b)}$ with $a^2+b^2=R^2$.
The action of the generators on the basis element $B^R_m(r,\varphi)$ is given by:
\begin{eqnarray}\label{eq43}
&&  \eta^{iso(2)}_{R}(X_1^{iso(2)})B^R_m  = -\frac{iR}{2}(B^R_{m+1}+B^R_{m-1})  \\ \nonumber
&&  \eta^{iso(2)}_{R}(X_2^{iso(2)})B^R_m  = \frac{R}{2}(-B^R_{m+1}+B^R_{m-1}) \\ \nonumber
&&  \eta^{iso(2)}_{r}(X_3^{iso(2)})B^R_m  = -imB^R_m  \nonumber
\end{eqnarray}

\begin{proposition}
$\eta^{iso(2)}_{R}$ is  the strong contraction of the family of representations $\left\{        ( \rho^{su(2)}_{(l,\epsilon_l^R)},V_{l}   )      \right\}_{l \in I_2}$ with respect to the contraction $su(2)\stackrel{t(\epsilon^R_l)}{\rightarrow} iso(2)$.
\end{proposition}
\textit{proof}.
Let $V$ be $\Omega_R$. We observe that for $\chi_{n,\epsilon_{n}^R}^{m}\in V_n$ we have

\begin{eqnarray}\label{eq44}
&&L(\chi_{n,\epsilon_{n}^R}^{m})(\theta,\varphi)= \lim_{l \longrightarrow \infty}\varphi_{n,l}(\chi_{n,\epsilon_{n}^R}^{m})(\theta,\varphi)=\lim_{l \longrightarrow \infty}\chi_{l,\epsilon_{l}^R}^{m}(\theta,\varphi)=\\ \nonumber
&& \lim_{l \longrightarrow \infty}Y_{l}^m(\frac{R}{l} \theta,\varphi)= (i)^mJ_m(R\theta)e^{-im\varphi}= B_{-m}^R(\theta,\varphi)
\end{eqnarray}
where $J_m$ is the Bessel function of order $m$ and we have used the asymptotic expansion of the spherical harmonics  (See page 229 of \cite{Vil}). Hence $(i)$ and $(ii)$ of definition \ref{def3} hold. Next for  any $\chi_{n,\epsilon_{n}^R}^{m}\in V_n$ and  $\chi_{k,\epsilon_{k}^R}^{s}\in V_k$ we have
\begin{eqnarray}\label{eq45}
&&\left\langle L(\chi_{n,\epsilon_{n}^R}^{m}),L(\chi_{k,\epsilon_{k}^R}^{s})\right\rangle=\left\langle   B_{-m}^R,  B_{-s}^R\right\rangle=\delta_{ms}\underbrace{=}_{\mbox{for } l\geq \max \left\{k,n\right\}}\\ \nonumber
&& \left\langle \varphi_{n,l}(\chi_{n,\epsilon_{n}^R}^{m})  ,\varphi_{k,l}(\chi_{k,\epsilon_{k}^R}^{s})  \right\rangle_{(l,\epsilon_l^R)}= \lim_{l \longrightarrow \infty} \left\langle \varphi_{n,l}(\chi_{n,\epsilon_{n}^R}^{m})  ,\varphi_{k,l}(\chi_{k,\epsilon_{k}^R}^{s})  \right\rangle_{(l,\epsilon_l^R)}
\end{eqnarray}
where we have used the fact that the embeddings $\varphi_{nl}$ preserve the inner product  and are compatible. This implies that $(iii)$ of definition \ref{def3} holds. We can also write (\ref{eq45}) in integral form  as:
\begin{eqnarray}
&& \int_{0}^{2\pi}\int_{0}^{\infty}(-i)^{-m}J_{-m}(R\theta)e^{-im\varphi}\overline{(-i)^{-s}J_{-s}(R\theta)e^{-is\varphi}}\theta d\theta d\varphi =\\ \nonumber
&& \int_{0}^{2\pi}\int_{0}^{\infty}\lim_{l \longrightarrow \infty}Y_{l}^m(\frac{R}{l} \theta,\varphi) \overline{\lim_{l \longrightarrow \infty}Y_{l}^s(\frac{R}{l} \theta,\varphi)}\theta d\theta d\varphi =\\ \nonumber
&& \lim_{l \longrightarrow \infty}\int_{0}^{2\pi}\int_{0}^{\frac{l\pi}{R}}Y_{l}^m(\frac{R}{l} \theta,\varphi) \overline{Y_{l}^s(\frac{R}{l} \theta,\varphi)}(2l+1) \sin{(\frac{R}{l}\theta)}d\theta d\varphi
\end{eqnarray}
For every $\chi_{n,\epsilon_{n}^R}^{m}\in V_n$ we have
\begin{eqnarray}\label{eq47}
&&\hspace{2.2cm}\lim_{l \longrightarrow \infty}  L\left(\rho^{su(2)}_{(l,\epsilon_l^R)}(t_{\epsilon_l^R}(X_{2}^{su(2)}))\varphi_{n,l}(\chi_{n,\epsilon_{n}^R}^m)\right)(\theta,\varphi)\underbrace{=}_{(\mbox{\ref{eq41}})} \\ \nonumber
&&\lim_{l \longrightarrow \infty} L\left( -\frac{Ri}{2l}( \sqrt{(l+m)(l-m+1)}\chi_{{l},\epsilon_{l}^R}^{m-1}+ \sqrt{(l-m)(l+m+1)}\chi_{{l},\epsilon_{l}^R}^{m+1})\right)(\theta,\varphi)=\\ \nonumber
&&\lim_{l \longrightarrow \infty}\lim_{k \longrightarrow \infty} \varphi_{lk}\left( -\frac{Ri}{2l}( \sqrt{(l+m)(l-m+1)}\chi_{{l},\epsilon_{l}^R}^{m-1}+ \sqrt{(l-m)(l+m+1)}\chi_{{l},\epsilon_{l}^R}^{m+1})\right)(\theta,\varphi)=\\ \nonumber
&&\lim_{l \longrightarrow \infty}\lim_{k \longrightarrow \infty}  -\frac{Ri}{2l}( \sqrt{(l+m)(l-m+1)}\chi_{{k},\epsilon_{k}^R}^{m-1}(\theta,\varphi)+  \sqrt{(l-m)(l+m+1)}\chi_{{k},\epsilon_{k}^R}^{m+1}(\theta,\varphi)) \underbrace{=}_{(\mbox{\ref{eq44}})}\\ \nonumber 
&& -\frac{iR}{2}( B_{-(m+1)}^R(\theta,\varphi)+B_{-(m-1)}^R(\theta,\varphi)  )\underbrace{=}_{(\mbox{\ref{eq43}})} \eta^{iso(2)}_{R}(\psi(X_2^{su(2)}))B^R_{-m}(\theta,\varphi)=\\ \nonumber
&& \eta^{iso(2)}_{R}(\psi(X_2^{su(2)}))L(\chi_{n,\epsilon_{n}^R}^{m})(\theta,\varphi)
\end{eqnarray}
Hence we have the desired pointwise convergence. We note that instead of taking both limits in (\ref{eq47}) we can set $k=l$ and take just one limit which is $l\longrightarrow \infty$ and we can rewrite (\ref{eq47}) in the following illuminating way:
\begin{eqnarray}
&&  \lim_{l \longrightarrow \infty}\rho^{su(2)}_{(l,\epsilon^R_l)}(t_{\epsilon_l^R}(X_2^{su(2)}))Y_{l}^m(\frac{R}{l} \theta,\varphi)  =\\ \nonumber
&& \lim_{l \longrightarrow \infty} \frac{R}{l}(-\cos{\varphi}\frac{l}{R}\frac{d}{d\theta}+\cot{(\frac{R}{l} \theta)}\sin{\varphi}\frac{d}{d\varphi} )Y_{l}^m(\frac{R}{l} \theta,\varphi)\\ \nonumber
&&  \left(-\cos{\varphi}\frac{d}{d\theta}+\frac{\sin{\varphi}}{\theta}\frac{d}{d\varphi}\right)(i)^mJ_m(R\theta)e^{-im\varphi}\\ \nonumber &&=\eta^{iso(2)}_{R}(\psi(X_2^{su(2)})) (i)^mJ_m(R\theta)e^{-im\varphi}
\end{eqnarray}
From the above we see that as we contract i.e., as we take the limit, the differential operators that act on the spherical harmonics along with the spherical harmonics themselves deform into differential operators that act on Bessel functions and into Bessel functions, respectively.
For convergence in norm we look at:
\begin{eqnarray}
&&  \left\|L\left(\rho^{su(2)}_{(l,\epsilon_l^R)}(t_{\epsilon_l^R}(X_{2}^{su(2)}))\varphi_{n,l}(\chi_{n,\epsilon_{n}^R}^m)\right)-\eta^{iso(2)}_{R}(\psi(X_2^{su(2)}))L(\chi_{n,\epsilon_{n}^R}^{m})\right\|^2 \hspace{5mm}\\ \nonumber
&&=   \left\|-\frac{iR}{2} \left(1-\frac{1}{l}\sqrt{(l-m)(l+m+1)}\right)B_{-(m+1)}^R\right\|^2+\\ \nonumber
&& \left\|\left(1-\frac{1}{l}\sqrt{(l+m)(l-m+1)}\right)B_{-(m-1)}^R\right\|^2 \underbrace{\longrightarrow}_{l \to \infty}0
\end{eqnarray}
where we have used (\ref{eq47}) and the fact that $\left\langle B_m^R,B_s^R\right\rangle=\delta_{ms}$.
Similarly we can prove these convergences for  $X_{1}^{su(2)},X_{3}^{su(2)}$.
Contraction of representations of $su(2)$ to those of $iso(2)$ was considered before  for example by  \.{I}n\"{o}n\"{u} and Wigner \cite{IW1} which used a different method.
\subsection{The contraction: $sl_2(\mathbb{R})\longrightarrow iso(2)$}
Set $\g=sl_2(\mathbb{R})$. The maps
\begin{eqnarray}
&& t_{\epsilon}:sl_2(\mathbb{R})\longrightarrow sl_2(\mathbb{R}) \\ \nonumber
&& t_{\epsilon}(X_1^{sl_2(\mathbb{R})})=\epsilon X_1^{sl_2(\mathbb{R})} \\ \nonumber
&& t_{\epsilon}(X_2^{sl_2(\mathbb{R})})=\epsilon X_2^{sl_2(\mathbb{R})}\\ \nonumber
&& t_{\epsilon}(X_3^{sl_2(\mathbb{R})})=- X_3^{sl_2(\mathbb{R})}
\end{eqnarray}
realize the contraction of the Lie algebra and lead to $ [X_3,X_1]_0=X_2, [X_3,X_2]_0= -X_1$
which induce the isomorphism
\begin{eqnarray}
&& \psi:\g_0\longrightarrow iso(2) \\ \nonumber
&& \psi(X_1^{sl_2(\mathbb{R})})= X_1^{iso(2)} \\ \nonumber
&& \psi(X_2^{sl_2(\mathbb{R})})=X_2^{iso(2)}  \\ \nonumber
&& \psi(X_3^{sl_2(\mathbb{R})})= X_3^{iso(2)}
\end{eqnarray}
For every $r \in \mathbb{R}$ the map
\begin{eqnarray}
&& \rho^{sl_2(\mathbb{R})}_{r\pm}:sl_2(\mathbb{R}) \longrightarrow gl(L^{2,\infty}([-\pi,\pi],dx)) \\ \nonumber
&& \rho^{sl_2(\mathbb{R})}_{r\pm}(X_1^{sl_2(\mathbb{R})})= -(ir \pm \frac{1}{2})\cos{x}+\sin{x}\frac{d}{dx}\\ \nonumber
&&   \rho^{sl_2(\mathbb{R})}_{r\pm}(X_2^{sl_2(\mathbb{R})})= (ir \pm \frac{1}{2})\sin{x}+\cos{x}\frac{d}{dx} \\ \nonumber
&& \rho^{sl_2(\mathbb{R})}_{r\pm}(X_3^{sl_2(\mathbb{R})})= -\frac{d}{dx}  \nonumber
\end{eqnarray}
is a SHIIR \cite{Vil}. This SHIIR  corresponds in the level of the group to the so  called principal unitary series representation  of $SU(1,1)\hspace{1mm}\cong SL_2(\mathbb{R})$ .\newline

For every $\epsilon \in I_1$ let $V_{\epsilon}$ be the inner product space $L^{2,\infty}([-\pi,\pi],dx)$.
For $\epsilon_i, \epsilon_j\in I_1$ such that $\epsilon_i \geq \epsilon_j$ we define the linear transformation   $\varphi_{\epsilon_i,\epsilon_j}:V_{\epsilon_i}\longrightarrow V_{\epsilon_j}$ to be the identity. $\varphi_{\epsilon_i,\epsilon_j}$ preserve the inner product and they are compatible.
\begin{proposition}
Let $r(\epsilon)=-\frac{r_2}{\epsilon}$. Then   $\rho^{iso(2)}_{(0,r_2)}$ regarded as a representation on $L^{2,\infty}([-\pi,\pi],dx)$  is  the strong contraction of the family of representations $\left\{        ( \rho^{sl_2(\mathbb{R})}_{r(\epsilon) +},V_{\epsilon}   )      \right\}_{\epsilon \in I_1}$ with respect to the contraction $sl_2(\mathbb{R})\stackrel{t(\epsilon)}{\rightarrow} iso(2)$.
\end{proposition}
The proof is similar to that of proposition \ref{prop1}.\newline
Contractions of representations of $sl_2(\mathbb{R})$ to those of  $iso(2)$  were considered before for example by  Celeghini and Tarlini \cite{{CeTa3}} which used a different method.
\subsection{The contraction: $sl_2(\mathbb{R})\longrightarrow \mathfrak{h}$}
\label{C9}
Set $\g=sl_2(\mathbb{R})$. The maps
\begin{eqnarray}
&& t_{\epsilon}:sl_2(\mathbb{R})\longrightarrow sl_2(\mathbb{R}) \\ \nonumber
&& t_{\epsilon}(X_1^{sl_2(\mathbb{R})})=-\epsilon X_1^{sl_2(\mathbb{R})} \\ \nonumber
&& t_{\epsilon}(X_2^{sl_2(\mathbb{R})}+X_3^{sl_2(\mathbb{R})})=X_2^{sl_2(\mathbb{R})}+X_3^{sl_2(\mathbb{R})}\\ \nonumber
&& t_{\epsilon}(X_2^{sl_2(\mathbb{R})})=\epsilon X_2^{sl_2(\mathbb{R})}
\end{eqnarray}
realize the contraction of the Lie algebra and lead to $[X_2,X_2+X_3]_0=X_1$
which induce the isomorphism
\begin{eqnarray}
&& \psi:\g_0\longrightarrow \mathfrak{h} \\ \nonumber
&& \psi(X_1^{sl_2(\mathbb{R})})= X_1^{\mathfrak{h}} \\ \nonumber
&& \psi(X_2^{sl_2(\mathbb{R})})=X_3^{\mathfrak{h}}  \\ \nonumber
&& \psi(X_2^{sl_2(\mathbb{R})}+X_3^{sl_2(\mathbb{R})})= X_2^{\mathfrak{h}}
\end{eqnarray}
For every $\epsilon \in I_1$ let $V_{\epsilon}$ be the inner product space $L^{2,\infty}_0([-\frac{ \pi}{ \epsilon},\frac{\pi}{\epsilon}],dx)$ and we define the function $\psi_{\epsilon}:[-\frac{ \pi}{ \epsilon},\frac{\pi}{\epsilon}]\longrightarrow [-\pi,\pi]$ by $\psi_{\epsilon}(x)=\epsilon x$.
We note that $P_{\epsilon}:L^{2,\infty}_0([-\pi,\pi],dx)\longrightarrow V_{\epsilon}$ which is given by $P_{\epsilon}(f)=f \circ \psi_{\epsilon}$, is  an isomorphism. Its inverse $P_{\epsilon}^{-1}$ is given by  $P_{\epsilon}^{-1}(f)=f \circ \psi^{-1}_{\epsilon}$.
For each  $\epsilon \in I_1$   we intertwine  $\rho^{sl_2(\mathbb{R})}_{r-}$ with $P_{\epsilon}$ to get the equivalent representation
\begin{eqnarray}
&&\rho^{sl_2(\mathbb{R})}_{(r-,\epsilon)}:sl_2(\mathbb{R})\longrightarrow gl(V_{\epsilon})\\ \nonumber
&&  \rho^{sl_2(\mathbb{R})}_{(r-,\epsilon)}(X_1^{sl_2(\mathbb{R})})=-(ir-\frac{1}{2})\cos{\epsilon x}+\sin{\epsilon x}\frac{1}{\epsilon}\frac{d}{dx}  \\ \nonumber
&& \rho^{sl_2(\mathbb{R})}_{(r-,\epsilon)}(X_2^{sl_2(\mathbb{R})})=(ir-\frac{1}{2})\sin{\epsilon x}+\cos{\epsilon x}\frac{1}{\epsilon}\frac{d}{dx} \\ \nonumber
&& \rho^{sl_2(\mathbb{R})}_{(r-,\epsilon)}(X_3^{sl_2(\mathbb{R})})=-\frac{1}{\epsilon}\frac{d}{dx}  \nonumber
\end{eqnarray}
For $\epsilon_i, \epsilon_j\in I_1$ such that $\epsilon_i \geq \epsilon_j$ we define the linear transformation   $\varphi_{\epsilon_i,\epsilon_j}:V_{\epsilon_i}\longrightarrow V_{\epsilon_j}$ to be the inclusion map that is defined as follows: for any $f\in V_{\epsilon_i}$, $\varphi_{\epsilon_i,\epsilon_j}(f)(x)=f(x)$ for $ x \in [-\frac{ \pi}{ \epsilon_i},\frac{\pi}{\epsilon_i}]$ and  $\varphi_{\epsilon_i,\epsilon_j}(f)(x)=0$ for $x \in [-\frac{ \pi}{ \epsilon_j},\frac{\pi}{\epsilon_j}]  \backslash [-\frac{ \pi}{ \epsilon_i},\frac{\pi}{\epsilon_i}]$.
These functions obviously preserve the inner product and are compatible.

\begin{proposition}
Let $r(\epsilon)=\frac{A}{\epsilon}$. Then   $\eta^{\mathfrak{h}}_A$ is  the strong contraction of the family of representations $\left\{        ( \rho^{sl_2(\mathbb{R})}_{(r(\epsilon)-,\epsilon)},V_{\epsilon}   )      \right\}_{\epsilon \in I_1}$ with respect to the contraction $sl_2(\mathbb{R})\stackrel{t(\epsilon)}{\rightarrow} \mathfrak{h}$, where we regard $\rho^{sl_2(\mathbb{R})}_{(r(\epsilon)-,\epsilon)}$ as a representation on $L^{2,\infty}_0([-\frac{ \pi}{ \epsilon},\frac{\pi}{\epsilon}],dx)$.
\end{proposition}
The proof is similar to that of proposition \ref{prop2}.

Contraction of the discrete series representation of $sl_2(\mathbb{R})$ to the representations  of $\mathfrak{h}$  was considered before by  Barut and  Girardello \cite{Bar} which used a different method.

\subsection{The contraction: $sl_2(\mathbb{R})\longrightarrow iso(1,1)$}
\label{C10}
Set $\g=sl_2(\mathbb{R})$. We will work with the basis
$$X^{sl_2(\mathbb{R})}=X_1^{sl_2(\mathbb{R})}-X_3^{sl_2(\mathbb{R})}, Y^{sl_2(\mathbb{R})}=X_1^{sl_2(\mathbb{R})}+X_3^{sl_2(\mathbb{R})}, H^{sl_2(\mathbb{R})}=2X_2^{sl_2(\mathbb{R})}.$$ In this basis the commutation relations are given by
\begin{eqnarray}
&& [H^{sl_2(\mathbb{R})},X^{sl_2(\mathbb{R})}]=2X^{sl_2(\mathbb{R})} \\ \nonumber
&& [H^{sl_2(\mathbb{R})},Y^{sl_2(\mathbb{R})}]=-2Y^{sl_2(\mathbb{R})}\\ \nonumber
&& [X^{sl_2(\mathbb{R})},Y^{sl_2(\mathbb{R})}]=H^{sl_2(\mathbb{R})}
\end{eqnarray}
The maps
\begin{eqnarray}
&& t_{\epsilon}:sl_2(\mathbb{R})\longrightarrow sl_2(\mathbb{R}) \\ \nonumber
&& t_{\epsilon}(X^{sl_2(\mathbb{R})})= X^{sl_2(\mathbb{R})} \\ \nonumber
&& t_{\epsilon}(Y^{sl_2(\mathbb{R})}-X^{sl_2(\mathbb{R})})=\epsilon (Y^{sl_2(\mathbb{R})}-X^{sl_2(\mathbb{R})})\\ \nonumber
&& t_{\epsilon}(H^{sl_2(\mathbb{R})})= \frac{1}{2}H^{sl_2(\mathbb{R})}
\end{eqnarray}
realize the contraction of the Lie algebra and lead to $[H,X]_0=X, [H,Y-X]_0=-(Y-X)$
which induce the isomorphism
\begin{eqnarray}
&& \psi:\g_0\longrightarrow iso(1,1) \\ \nonumber
&& \psi(X^{sl_2(\mathbb{R})})= X_1^{iso(1,1)} \\ \nonumber
&& \psi(Y^{sl_2(\mathbb{R})}-X^{sl_2(\mathbb{R})})=X_2^{iso(1,1)}  \\ \nonumber
&& \psi(H^{sl_2(\mathbb{R})})= X_3^{iso(1,1)}
\end{eqnarray}
Before we contract the representations of the Lie algebra, we give the SHIIR of $\mathfrak{g}(-1)=iso(1,1)$ in a different realization which better suits our needs.  We define a linear isometry $\tau :L^{2,\infty}_c(\mathbb{R},dx) \longrightarrow L_c^{2,\infty}((0,\infty),\frac{dx}{x})$ by $\tau (f)=f\circ\ln$. Intertwining $\rho_{(a,b)}^{iso(1,1)}$ with $\tau$ we obtain the following SHIIR
\begin{eqnarray}
&& \bar{\rho}^{iso(1,1)}_{(a,b)}:iso(1,1)\longrightarrow gl(L_c^{2,\infty}((0,\infty),\frac{dx}{x})) \\ \nonumber
&& \bar{\rho}^{iso(1,1)}_{(a,b)}(X_1^{iso(1,1)})=  iax\\ \nonumber
&&   \bar{\rho}^{iso(1,1)}_{(a,b)}(X_2^{iso(1,1)})= -ib\frac{1}{x} \\ \nonumber
&& \bar{\rho}^{iso(1,1)}_{(a,b)}(X_3^{iso(1,1)})= x\frac{d}{dx}  \nonumber
\end{eqnarray}
In fact the representations $\left\{\bar{\rho}^{iso(1,1)}_{(\pm 1,b)}|b\in \mathbb{R}\right\}$ exhaust all the SHIIR of $iso(1,1)$.
For every $n \in \mathbb{N}_0$ the map
\begin{eqnarray}
&& \bar{\rho}^{sl_2(\mathbb{R})}_{n\pm}:sl_2(\mathbb{R}) \longrightarrow gl(L_c^{2,\infty}((0,\infty),\frac{dx}{x})) \\ \nonumber
&& \bar{\rho}^{sl_2(\mathbb{R})}_{n\pm}(X^{sl_2(\mathbb{R})})= \pm ix\\ \nonumber
&&   \bar{\rho}^{sl_2(\mathbb{R})}_{n\pm}(Y^{sl_2(\mathbb{R})})= -i\frac{n^2-1}{4}\frac{1}{x}+ix\frac{d^2}{dx^2} \\ \nonumber
&& \bar{\rho}^{sl_2(\mathbb{R})}_{n\pm}(H^{sl_2(\mathbb{R})})= 2x\frac{d}{dx}  \nonumber
\end{eqnarray}
is a SHIIR. This realization is known as the Kirillov model and it was recently found at \cite{Baruch}. These representations corresponds in the level of the group to the so called  discrete series representation  of $SU(1,1)\hspace{1mm}\cong SL_2(\mathbb{R})$ .\newline

For every $n \in \mathbb{N}_0$ let $V_n$ be the inner product space $L_c^{2,\infty}((0,\infty),\frac{dx}{x})$.
For $m, n\in \mathbb{N}_0$ such that $n \geq m$ we define the linear transformation   $\varphi_{m,n}:V_{m}\longrightarrow V_{n}$ to be the identity. These functions preserve the inner product and are compatible.

\begin{proposition}
Let $\epsilon_n=\frac{4b}{n^2}$. Then   $\bar{\rho}^{iso(1,1)}_{(\pm 1,b)}$ is  the strong contraction of the family of representations $\left\{        ( \bar{\rho}^{sl_2(\mathbb{R})}_{n\pm},V_{\epsilon_n}   )      \right\}_{\epsilon \in I_2}$ with respect to the contraction $sl_2(\mathbb{R})\stackrel{t_{\epsilon_n}}{\rightarrow} iso(1,1)$.
\end{proposition}

\textit{proof}.
Let $V$ be $L_c^{2,\infty}((0,\infty),\frac{dx}{x})$. Since the  embeddings $\varphi_{mn}$ are the identity map they  are obviously compatible and satisfy conditions $(i), (ii)$ and $(iii)$ in definition \ref{def3} and also $L(f)=f$ for every $f(x) \in L_c^{2,\infty}((0,\infty),\frac{dx}{x})$. For condition $(iv)$ we observe that for  $f \in V_{n}=L_c^{2,\infty}((0,\infty),\frac{dx}{x})$ we have:
\begin{eqnarray}
&&\lim_{n \longrightarrow \infty}  L(\bar{\rho}^{sl_2(\mathbb{R})}_{n\pm}(t_n(H^{sl_2(\mathbb{R})}))(f))(x)= \\ \nonumber
&&\lim_{n \longrightarrow \infty}  \bar{\rho}^{sl_2(\mathbb{R})}_{n\pm}(t_n(H^{sl_2(\mathbb{R})}))(f)(x) =  \lim_{n \longrightarrow \infty}  xf'(x)=xf'(x)=x\frac{d}{dx}L(f)(x) =\\ \nonumber
&&\bar{\rho}^{iso(1,1)}_{(\pm 1,b)}(\psi(H^{sl_2(\mathbb{R})}))L(f)(x)
\label{eq}
\end{eqnarray}
\begin{eqnarray}
&&\lim_{n \longrightarrow \infty}  L(\bar{\rho}^{sl_2(\mathbb{R})}_{n\pm}(t_n(Y^{sl_2(\mathbb{R})}-X^{sl_2(\mathbb{R})}))(f))(x) =\\ \nonumber
&&\lim_{n \longrightarrow \infty}  \bar{\rho}^{sl_2(\mathbb{R})}_{n\pm}(t_n(Y^{sl_2(\mathbb{R})}-X^{sl_2(\mathbb{R})}))(f)(x)= \\ \nonumber
&&\lim_{n \longrightarrow \infty} \epsilon_n(-i\frac{n^2-1}{4}\frac{1}{x}+ix\frac{d^2}{dx^2}-\pm ix  ) f(x) \\ \nonumber
&&\lim_{n \longrightarrow \infty} \frac{4b}{n^2}(-i\frac{n^2-1}{4}\frac{1}{x}+ix\frac{d^2}{dx^2}-\pm ix  ) f(x) =-ib\frac{1}{x}f(x)=\\ \nonumber         &&=-ib\frac{1}{x}L(f)(x)=\bar{\rho}^{iso(1,1)}_{(\pm 1,b)}(\psi (Y^{sl_2(\mathbb{R})}-X^{sl_2(\mathbb{R})}))L(f)(x)
\end{eqnarray}
\begin{eqnarray}
&&\lim_{n \longrightarrow \infty}  L(\bar{\rho}^{sl_2(\mathbb{R})}_{n\pm}(t_n(X^{sl_2(\mathbb{R})}))(f))(x) = \\ \nonumber
&&\lim_{n \longrightarrow \infty}  \bar{\rho}^{sl_2(\mathbb{R})}_{n\pm}(t_n(X^{sl_2(\mathbb{R})}))(f)(x)=\lim_{n \longrightarrow \infty} (\pm ix  ) f(x) =\pm ix L(f)(x)=\\ \nonumber
&&\bar{\rho}^{iso(1,1)}_{(\pm 1,b)}(\psi (X^{sl_2(\mathbb{R})}))L(f)(x)
\end{eqnarray}
The above pointwise limit can be shown to be a uniform limit due to the compactness of the support of $f$ and this implies that we also have the desired convergence in norm.

Contractions of representations of $sl_2(\mathbb{R})$ to those of $iso(1,1)$   were considered before for example by  Celeghini and Tarlini \cite{{CeTa3}} and by Barut and Fornsdal \cite{Bar2} which used  different methods.\newline

\section{Discussion}

\subsection{Convergence of matrix elements versus convergence of differential operators}
The early work of \.{I}n\"{o}n\"{u} and Wigner \cite{IW1} presented two approaches to contraction of Lie algebra representations.
In the first they started with a representation of $\g$ which was given by differential operators on some function space. Then they deformed the representation by conjugating it in a way which is dependent on the contraction parameter and leads, under a certain limit, to a representation of $\g_0$. This approach was demonstrated in the contraction of the two-dimensional affine Lie algebra to the abelian Lie algebra (\cite{IW1} p. 514-515).
In the second approach they started with a sequence of representations of $\g$ such that the matrix elements of the sequence converge to the matrix elements of some fixed representation of $\g_0$. The contraction of the representations of the Lie algebra of $so(3)$ to those of $iso(2)$ was given as an example (\cite{IW1} p. 516-517).\newline

A question that arises naturally is: Can any contraction of Lie algebra representations  be realized in both of these approaches ?
Since strong contraction of Lie algebra representations is a generalization of the first approach, this paper gives the answer in the three-dimensional case.  As we have shown, in the three-dimensional case any SHIIR of the limit Lie algebra, $\g_0$ is a  contraction of Lie algebra representations that can be realized as in the first approach of \.{I}n\"{o}n\"{u} and Wigner, i.e., as a (strong) contraction of differential operators on some space of functions. As to the method of convergence of matrix elements, given a contraction (see definition 4 in the appendix) or a strong contraction  of Lie algebra representations and  compatible (see appendix) bases of the representation spaces,  by proposition \ref{Prop14} in the appendix we obtain also a contraction by means of convergence of matrix elements. We emphasize that the existence of compatible bases is not assured. We believe that the strong contractions of representations that was given here in the cases:
\begin{itemize}
	\item $iso(2)\longrightarrow \mathfrak{h}$
	\item $sl_2(\mathbb{R})\longrightarrow \mathfrak{h}$		
\end{itemize}
do not have compatible bases and therefore cannot be obtained by the method of contraction of matrix elements.

\subsection{The family of embeddings and the realization of the limit space}
In our construction the limit representation space is being realized through pointwise limit of functions  under a certain family of compatible embeddings. These embeddings were implicit in the work of \.{I}n\"{o}n\"{u}, Wigner and others but for our construction they are crucial ingredients. Moreover, as it is proved in the appendix, such a family of compatible embeddings gives rise to a direct limit space. In particular, it follows from our  work that every SHIIR of the limit three-dimensional Lie algebra is a direct limit space. \newline

The necessity of  these embeddings puts some limitations on the realizations of the representations that we can use. It turns out that only certain realizations are  suitable for strong contractions. Occasionally it is very difficult to find  these particular realizations.For example in the strong contractions of the discrete series representations of $sl_2(\mathbb{R})$ to the  irreducible representations of the Poincare groups $iso(1,1)$
(section \ref{C10})  we have used  the highly non trivial Kirillov model for the discrete series representations of $sl_2(\mathbb{R})$ whose full description was only obtained lately \cite{Baruch}  by the second author. We have tried several different standard realizations of the discrete series for this contraction but the only one that worked was the the realization of the Kirillov model.\newline
 Another difficulty is that the same group representation induces  more than one representation of the associated Lie algebra. Usually one takes the dense subspace  of compactly supported smooth functions inside the representation space of the Lie group, but there are several exceptions. For example in the strong contraction of representations for the contractions $ iso(2)\longrightarrow \mathfrak{h}$ and $sl_2(\mathbb{R})\longrightarrow \mathfrak{h}$ (sections \ref{C2} and \ref{C9} respectively) instead of working with the subspaces $L^{2,\infty}([-\frac{\pi}{\epsilon},\frac{\pi}{\epsilon}],dx)$  we have used the subspaces $L^{2,\infty}_0([-\frac{\pi}{\epsilon},\frac{\pi}{\epsilon}],dx)$ as representation spaces for $ iso(2)$ and $sl_2(\mathbb{R})$. The fact that  all the derivatives of any  function in $L^{2,\infty}_0([-\frac{\pi}{\epsilon},\frac{\pi}{\epsilon}],dx)$ vanish at $\pm\frac{\pi}{\epsilon}$ allowed us to consider the extension of each function by zero where it was not defined previously  as an embedding. These maps obviously are not embeddings for the spaces $L^{2,\infty}([-\frac{\pi}{\epsilon},\frac{\pi}{\epsilon}],dx)$. \newline

Another problem is that in many cases we needed to intertwine  our representations before contracting them. The intertwiner was dependent on the contraction parameter and was difficult to find. For example in section \ref{C1} the  function $\nu_{\epsilon}(x)=e^{-\epsilon x}$  served as an intertwiner  from the representations of $\mathfrak{ea}$ on $L^{2\infty}_c(\mathbb{R}^+,\frac{dx}{x})$ to  equivalent representations on  $L^{2\infty}_c(\mathbb{R},dx)$.

\subsection{New results}\label{se}
The strong contractions of representations in the cases:
\begin{itemize}
	\item $\mathfrak{ea}\longrightarrow \mathfrak{h}$
	\item $iso(2)\longrightarrow \mathfrak{h}$
	\item $\mathfrak{g}(\lambda)_{\lambda \neq 1}\longrightarrow \mathfrak{h}$
	\item $\mathfrak{l}(\lambda)_{\lambda\neq 0}\longrightarrow \mathfrak{h}$
	\item	$\mathfrak{c}\longrightarrow \mathfrak{h}$
	\item $\mathfrak{c}\longrightarrow \mathfrak{g}(1)$
	\item $sl_2(\mathbb{R})\longrightarrow \mathfrak{h}$		
\end{itemize}
 are  new. Barut and Girardello \cite{Bar} gave a contraction of the Lie algebra representations for $sl_2(\mathbb{R})\longrightarrow \mathfrak{h}$ but they contract the discrete series representations of $sl_2(\mathbb{R})$ while we contract the continuous series representations. This scenario seems to be a first example for contraction of two inequivalent families of representations of the same Lie algebra that contract to the same representation. 

\subsection{Related works}

Contraction of Lie algebra representations in which the representations are realized by differential operators that act on some functions spaces was considered systematically by, e.g., \cite{Izm1,Izm2,Kal,Izm3,Izm4,Pog}. Their main objective was to look at separation of variables under contraction. While there are some similarities between their work and some examples that we studied, they did not consider the new examples that were mentioned in section 6.3. Moreover our work emphasizes the direct limit structure including the compatible embeddings and the inner product structure.

\ack
\begin{itemize}
	\item AM is grateful to Prof. Weimar-Woods for a helpful discussion and to Prof. Wei-Min Zhang for his very kind hospitality in
Tainan. The work of AM was partly supported by grant HUA97-12-02-161
at NCKU.
\item The research of the 2nd author was supported by
the center of excellence of the Israel Science Foundation
grant no. 1691/10.
\item JLB thanks the Department of Physics, Technion, for its warm
hospitality and support during visits while this work was
being carried out, and the FRAP-PSC-CUNY for some support.
\end{itemize}

\appendix

\section{Contraction of Lie algebra representations as direct limit}

In this appendix we give  a  definition for contraction of Lie algebra representations in terms of direct limit \cite{Mac}. We discuss the relation between three types of contractions: strong contraction, general contraction in terms of direct limit and contraction as a limit of matrix elements. The first to consider the question of contraction of Lie algebra representations  were \.{I}n\"{o}n\"{u} and Wigner \cite{IW1}; they describe, in some particular examples, how to build faithful representations of $\g_0$ from a family of representations of $\g$ by some limiting procedure. After the pioneering work of \.{I}n\"{o}n\"{u} and Wigner there have been several attempts to give a  general procedure for the contraction of Lie algebras representations e.g ., \cite{Wei2,CeTa1,CeTa3,Sal,Cat,CeTa2,CW,Pat}, but up to now there does not seem to be a generally accepted definition. All the known examples fit within our scheme of contraction of representations using direct limit as explained below.

\subsection{Direct limit of inner product spaces}

Let $I=(I,\prec)$ be a totally ordered set. Suppose that for every $i\in I$ we are given an inner product space $(V_i, \left\langle \_,\_\right\rangle_i )$ and for every $i,j \in I$ such that $i\prec j$  we have a linear map $\varphi_{ij}:V_i\longrightarrow V_j$ which preserves the inner product. If, in addition, for every $i,j,k \in I$ such that $i\prec j \prec k$ we have $\varphi_{ik}=\varphi_{jk} \circ \varphi_{ij}$,	and for all $i \in I$  $\varphi_{ii}$ is the identity operator on $V_i$, we call
such a collection of inner product spaces and linear maps a directed system of inner product spaces over $I$. We will call it for short a directed system and denote it by  $\left\{V_i, \varphi_{ij}, I\right\}$.
Given a directed system $\left\{V_i, \varphi_{ij}, I\right\}$ we denote its direct limit  by  $V^{\infty}$.
We will use the following construction of the direct limit:  let $\left\{V_i, \varphi_{ij}, I\right\}$ be a directed system. Let $X$ be the disjoint union of the $V_i$ ($X=\coprod_{i\in I}V_i$). To indicate that $x \in \coprod_{i\in I}V_i$ belongs to $V_i$ we will write $x^i$ instead of $x \in V_i$.  We define an equivalence relation on $X$ by: $x^i\sim y^j$ if there exists   $k\in I$ such that $i \prec k$, $j \prec k$ and $\varphi_{ik}(x^i)=\varphi_{jk}(y^j)$. We denote the equivalence class of $x$ by $[x]$ and the collection of the equivalence classes by $V^{\infty}=X/\sim$. We give $V^{\infty}$ a structure of inner product space by defining the multiplication of $[x^i]$ by a scalar $\alpha$ to be $[\alpha x^i]$, the sum of $[x^i]$ and $[y^j]$ is defined to be  $[\varphi_{ik}(x^i)+\varphi_{jk}(y^j)]$ and the inner product $<[x^i],[y^j]>$ is given by  $<\varphi_{ik}(x^i),\varphi_{jk}(y^j)>_k$  where $k\in I$ is such that $i \prec k$, $j \prec k$. $V^{\infty}$ as constructed here is the direct limit of the directed system $\left\{V_i, \varphi_{ij}, I\right\}$, i.e.
\begin{eqnarray}
V^{\infty}=\coprod_{i\in I}V_i/\sim
\end{eqnarray}
 We remark that for each $i\in I$ we have the natural morphism $\varphi_i:V_i\longrightarrow V^{\infty} $ which  sends $x$ to its equivalence class, $[x]$. This map is obviously linear and preserves the inner product and hence it is an embedding of $V_i$ in $V^{\infty}$. The maps $\left\{\varphi_i\right\}_{i \in I}$ are compatible with the directed system in the following sense: For every $i,j \in I$ such that $i \prec j$, $\varphi_i=\varphi_j\circ \varphi_{ij}$.
\begin{example}
Let $I$ be the totally ordered set which is $\mathbb{N}$ with the usual order of natural numbers. Let $\left\{V_i, \varphi_{ij}, I\right\}$ be the directed system of inner product spaces over $I$ which is defined as follows: For every $n \in \mathbb{N} $, $V_n=\mathbb{C}^n$ with the standard inner product. And for $m\leq n$ we have $\varphi_{mn}\left((x_1,x_2,...,x_m)\right)=(x_1,x_2,...,x_m,0,0,...,0)$. In this case the direct limit is  the inner product space of all infinite sequences with finite numbers of non zero entries and with the standard inner product. \newline
In terms of the above construction the elements of the direct limit are given by
$$[(x_1,x_2,...,x_m)]=\left\{(x_1,x_2,...,x_m),(x_1,x_2,...,x_m,0),(x_1,x_2,...,x_m,0,0),...\right\}  \; $$
where $x_m\neq 0 $. The inner product of $[(x_1,x_2,...,x_m)], [(y_1,y_2,...,y_n)] \in V^{\infty} $ where $m\leq n$ is given by $\left\langle [(x_1,x_2,...,x_m)], [(y_1,y_2,...,y_n)]\right\rangle=\sum_{i=1}^{m} \bar{x}_i y_i$.
We identify  $[(x_1,x_2,...,x_m)]$ with  $(x_1,x_2,...,x_m,0,0,......).$
\end{example} 

\subsection{Contraction as direct limit}

In the case of sequential contraction and a countable ordered set let $I_2$ denote the totally ordered set which consists of the natural numbers, $\mathbb{N}$ and the usual order of natural numbers, i.e. $m \prec n$ $\Longleftrightarrow$ $m\leq n$.\newline

\begin{definition}
Suppose $\g\stackrel{t_n}{\rightarrow} \g_{\infty}$. Let $\left\{V_{i}, \varphi_{ij}, I_2\right\}$  be a directed system of inner product spaces  and $\left\{(\rho_{i},V_{i})\right\}_{i \in I_2}$ a sequence of representations of $\g$. A representation $\eta:\g_{\infty}\longrightarrow gl(W)$ is called a contraction of the sequence of representations $\left\{(\rho_{i},V_{i})\right\}_{i \in I_2}$ with respect to
the contraction $\g\stackrel{t_n}{\rightarrow} \g_{\infty}$ and we denote it by $\rho_n\stackrel{t_n}{\rightarrow} \eta$ if
\begin{enumerate}
	\item For every $[v^m]\in V^{\infty}$, $X\in U$, the limit  $$\rho_{\infty}(X)[v^m]\equiv \lim_{n\longrightarrow \infty}[\rho_n(t_{\epsilon_n}(X))\varphi_{mn}(v^{m})]$$ exists
  \item There exists a linear invertible transformation $K:V^{\infty} \longrightarrow W$ such that for every $[v]\in V^{\infty}$, $X\in U$ we have  $\rho_{\infty}(X)[v]=K^{-1} \eta(X)  K ([v]) $
\end{enumerate}
 \label{def4}
\end{definition}

\textbf{Remark 3.}
\begin{itemize}
 \item $\rho_{\infty}(X)[v^{m}]$ does not depend on the representative $v^{m}$, since if $[v^{k}]=[v^{m}]$ and  $m \geq k$ so $v^m=\varphi_{km}(v^k)$ $\Longrightarrow$ $[\rho_{n}(t_{\epsilon_n}(X))\varphi_{mn}(v^m)]=[\rho_{n}(t_{\epsilon_n}(X))\varphi_{mn}(\varphi_{km}(v^k))]=[\rho_{n}(t_{\epsilon_n}(X))\varphi_{kn}(v^k)]$.
	\item  $V^{\infty}$ is an inner product space and hence also a metric space so the notion of limit of elements of $V$ is defined.
	\item The existence of the limit representation $\eta$ in the definition is not essential. We could demand instead of condition (2) that the mapping $X\longmapsto \rho_{\infty}(X)$  will be a representation of $\g_{\infty}$ on the space $V^{\infty}$. In practice it is easier to find the intertwiner $K$.
\end{itemize}

In the continuous case i.e., when $\g\stackrel{t(\epsilon)}{\rightarrow} \g_0$ and we have a family of representations of $\g$  instead of $I_2$, we work with  $I_1$ which we define to be  the totally ordered set which consists of the interval $(0,1]$ and the order relation $\prec$ defined by: $x \prec y$ $\Longleftrightarrow$ $x\geq y$. Then we have a similar definition to the sequential case in which  we change the above limit to $\epsilon \longrightarrow 0^+$.
Obviously every contraction of a sequence of representations can be turned into a contraction of a family of representations.\newline

We remark that in this paper when we say that $\rho:\g\longrightarrow gl(W)$ is a representation of the Lie algebra $\g$ and $B$ is  a basis of $W$,  we mean that every vector $w\in W$ can be written as a finite linear combination of elements of $B$ and not in the sense of a basis of a Hilbert space. Our choice for linear spaces with an algebraic basis i.e., every vector is a finite linear combination of the basis elements, is motivated by  the Harish-Chandra theory of $(g,K)$-modules  \cite{Wal} in which the representation spaces have this property. We plan to address the question of completion to Hilbert space  in a future paper where we also deal with contraction of the representations of  Lie groups.

\begin{proposition}
In the notations of the  definition \ref{def4}  when $\rho_{n}\stackrel{t_n}{\rightarrow} \eta$, for every $[v^{m}], [v^{k}] \in V^{\infty}$ and $X\in U$
\begin{eqnarray}
\left\langle[v^m] ,\rho_{\infty}(X) [v^k]\right\rangle=\lim_{n\longrightarrow \infty} \left\langle \varphi_{mn}(v^{m}),\rho_{n}(t_{\epsilon_n}(X))\varphi_{kn}(v^k)\right\rangle_{n}.
\end{eqnarray}
\label{Prop11}
\end{proposition}

\textbf{proof.}
\begin{eqnarray}
&& \left\langle[v^m] ,\rho_{\infty}(X) [v^k]\right\rangle=\left\langle[v^{m}] ,\lim_{n\longrightarrow \infty}[\rho_{n}(t_{\epsilon_n}(X))\varphi_{kn}(v^{k})]\right\rangle= \\ \nonumber
&& \lim_{n\longrightarrow \infty}\left\langle[v^{m}] ,[\rho_{n}(t_{\epsilon_n}(X))\varphi_{kn}(v^{k})]\right\rangle= \\ \nonumber
&&\lim_{n\longrightarrow \infty}\left\langle \varphi_{mn}(v^{m}) ,\rho_{n}(t_{\epsilon_n}(X))\varphi_{kn}(v^{k})\right\rangle_n \nonumber
\label{eq:77}
\end{eqnarray}
where the second equality is true since the inner product is continuous and the last equality follows from the definition of the inner product  in $V^{\infty}$ and the fact that $n$ is very large.\newline
We note that if in addition $K$ is unitary then we  have
\begin{eqnarray}
&&\left\langle K[v^{m}] ,\eta(X) K[v^{k}]\right\rangle_W= \left\langle[v^{m}] ,\rho_{\infty}(X) [v^{k}]\right\rangle=\\ \nonumber && \lim_{n\longrightarrow \infty}\left\langle \varphi_{mn}(v^{m}) ,\rho_{n}(t_{\epsilon_n}(X))\varphi_{kn}(v^{k})\right\rangle_n \nonumber
\end{eqnarray}
where $\left\langle \_, \_\right\rangle_W$ stands for the inner product on $W$.

\begin{proposition}
Suppose $\g\stackrel{t_n}{\rightarrow} \g_{\infty}$. Let $\left\{V_{n}, \varphi_{mn}, I_2\right\}$  be a directed system of inner product spaces and $\left\{(\rho_{n},V_{n})\right\}_{n \in I_2}$ a sequence of representations of $\g$. Suppose that $\eta:\g_{\infty}\longrightarrow gl(W)$ is a representation such that $W$ has a countable orthonormal basis. Assume that for every $n \in I_2$ there is a linear transformation $\tau_{n}:V_{n}\longrightarrow W$ that preserves the inner product,  such that:
\begin{enumerate}
	\item For every $k \leq m$, $\tau_k=\tau_m  \circ\varphi_{km} $ \label{Prop2c1}
	\item For every $w\in W$ there is  some $[v^m]\in V^{\infty} $, such that  $\tau_{m}(v^m)=w$ \label{Prop2c2}
		\item For every $X\in U$, $[v^{m}],[v^{k}] \in V^{\infty}$ $$\lim_{n\longrightarrow \infty} \left\langle \varphi_{mn}(v^{m}),\rho_{n}(t_{\epsilon_n}(X))\varphi_{kn}(v^{k})\right\rangle_{n} = \left\langle \tau_{m}(v^{m}),\eta(X)\tau_{k}(v^{k})\right\rangle_{W}$$\label{Prop2c3}
			\item For every $X \in U$ and every  $[v^{m_0}]\in V^{\infty}$ $$\dim span \left\{   \bigcup_{ m_0 \leq n}\tau_{n}\rho_{\epsilon_n}(X)\varphi_{m_0n}(v^{m_0}) \right\} < \infty$$ \label{Prop2c4}
\end{enumerate}			
Then  $(\eta,W)$ is a contraction of the sequence of representations $\left\{(\rho_{n},V_{n})\right\}_{n \in I_2}$ with respect to
the contraction $\g\stackrel{t(n)}{\rightarrow} \g_{\infty}$.
\label{Prop12}	
\end{proposition}

\textbf{proof.} Let $V^{\infty}$ be the direct limit of $\left\{V_{n}, \varphi_{mn}, I_2\right\}$. We define a map $K:V^{\infty}\longrightarrow W$ by $K([v^{m} ])=\tau_{m}(v^{m})$. It follows from \textit{(i), (ii)} and \textit{(iii)} that $K$ is well defined linear isometry.
In addition for any $X \in U$, $[v^{m}],[v^{k}] \in V^{\infty}$ we have :
\begin{eqnarray}
&&  \lim_{n\longrightarrow \infty}\left\langle[v^m] ,[\rho_{\epsilon_n}(t_{\epsilon_n}(X))\varphi_{kn}(v^{k})]\right\rangle
 = \\ \nonumber
 &&\lim_{n\longrightarrow \infty} \left\langle \varphi_{mn}(v^{m}),\rho_{\epsilon_n}(t_{\epsilon_n}(X))\varphi_{kn}(v^{k})\right\rangle_{n}  \\
&& \underbrace{=}_{(\ref{Prop2c3})} \left\langle \tau_{m}(v^{m}),\eta(X)\tau_{k}(v^{k})\right\rangle_{W}= \left\langle K([v^{m}]),\eta(X)K([v^{k}])\right\rangle_W  \nonumber
\label{eq:77}
\end{eqnarray}
And since $K^{-1}$ is an isometry  we get
\begin{eqnarray}
&& \lim_{n\longrightarrow \infty}\left\langle[v^m] ,[\rho_{\epsilon_n}(t_{\epsilon_n}(X))\varphi_{kn}(v^{k})]\right\rangle=\left\langle [v^{m}],K^{-1}\eta(X)K([v^{k}])\right\rangle
\end{eqnarray}
i.e., $ \left\{[\rho_{\epsilon_n}(t_{\epsilon_n}(X))\varphi_{kn}(v^{k})]\right\}_{n\in \mathbb{N}}$  weakly converges to $K^{-1}\eta(X)K([v^{k}])$ when $n$ goes to infinity. From \textit{(iv)} and  the  fact that weak convergence in finite dimensional spaces is equivalent to convergence in norm it follows that for every $X \in U$ and $[v^{m}]\in V^{\infty}$
$ \left\{[\rho_{n}(t_{\epsilon_n}(X))\varphi_{mn}(v^{m})]\right\}_{n\in \mathbb{N}}$  converges  in norm to $K^{-1}\eta(X)K([v^{m}])$ as $n$ goes to infinity.
Hence for every $[v^{m}]\in V^{\infty}$, $X\in U$ the limit  $\rho_{\infty}(X)[v^{m}]$ exists and we have  $\rho_{\infty}(X)[v^{m}]=K^{-1} \eta(X)  K ([v^m])$.
So we have proved that $\rho_{\epsilon}\stackrel{t(\epsilon)}{\rightarrow} \eta$.\newline

Proposition  \ref{Prop12} actually shows that convergence of matrix elements as it usually appears in the context of contraction of representations implies contraction of Lie algebra representations according to the given definition in this work.\newline
A sequence of bases,  $\left\{B_n \right\}_{n\in I_2}$ will be called compatible if for every $m\geq n$ we have $\varphi_{nm}(B_n)\subseteq B_m$ and for each element in these bases  condition \textit{(iv)} of proposition \ref{Prop12} holds.

\begin{proposition}
If $\eta:\g_{\infty}\longrightarrow gl(V)$ is  the strong contraction of the sequence of representations $\left\{(\rho_{n},V_{n})\right\}_{n \in  \mathbb{N}}$ with respect to the contraction $\g\stackrel{t_n}{\rightarrow} \g_{\infty}$ then it is also a contraction in the sense of definition \ref{def4}.
\label{Prop13}
\end{proposition}

\textbf{proof.}
Define a map  $K$ from the direct limit $V^{\infty}$ to $V$ by $K([f])=L(f)$ for every $[f] \in V^{\infty}$. One can show that $K$ is the desired intertwiner in the sense of definition \ref{def4}.

\begin{proposition}
Let $\eta:\g_{\infty}\longrightarrow gl(V)$ be  the strong contraction of the  sequence of representations $\left\{(\rho_{n},V_{n})\right\}_{n \in  \mathbb{N}}$ with respect to the contraction  $\g\stackrel{t_n}{\rightarrow}\g_{\infty}$. For any $n\in \mathbb{N}$ let $B_n=\left\{b_i^{n}|i\in I(n)\right\}$, where $I(n)$ is some set of indices,  be a basis for $V_n$. Suppose that  the sequence    $\left\{B_n\right\}_{ n\in \mathbb{N}}$ is compatible, i.e., for every $b_i^{m}\in B_{m}$  $\varphi_{mn}(b_i^{m})=b_i^{n}\in B(n)$. Then $B=\left\{L(b_i^{n})|i \in \mathbb{N}, i\in I(n))\right\}$ is a basis for $V$ and we have the following convergence of matrix elements: for any $b_i^{m}\in B(m)$, $b_j^{n}\in B(n)$ and any $X \in U $we have
\begin{eqnarray}
&&  \lim_{k\longrightarrow \infty}\left\langle \varphi_{mk}(b_i^{m}) ,\rho_{k}(t(\epsilon_k)X)\varphi_{nk}(b_j^{n})\right\rangle_{k}
 = \left\langle L(b_i^{m}) ,\eta(X)L(b_j^{n})\right\rangle
\end{eqnarray}
\label{Prop14}
\end{proposition}

The proof follows from propositions \ref{Prop11} and \ref{Prop13}.\newline
A similar statement for contraction of representations  with compatible bases also holds.



\section*{References}
\begin{thebibliography}{10}
\bibitem {Wei3} E. Weimar-Woods,  Rev. Math. Phys. \textbf{12}, 1505 (2000).
\bibitem {IW1} E. \.In\"on\"u, and E. P. Wigner, Proc. Nat. Acad. Sci. U.S \textbf{39}, 510 (1953).

\bibitem {IW2} E. \.In\"on\"u and E. P. Wigner,  Proc. Nat. Acad. Sci. U.S \textbf{40}, 119 (1954).

\bibitem {Wei2} E. Weimar-Woods,  J. Math. Phys. \textbf{32}, 2660 (1991).

\bibitem {CeTa1} E. Celeghini and M. Tarlini,  Il Nuovo Cimento B \textbf{61}, 265 (1981).

\bibitem {Bar} Barut A.O. and L. Girardello,   Commun. Math. Phys. \textbf{21}, 41 (1971).

\bibitem {Bar2} Barut A.O. and C. Fronsdal,    Proceedings of the Royal Society of London. Series A, Mathematical and Physical Sciences,  \textbf{287} (1411), 532 (1965).

\bibitem {Gil}Robert Gilmore,  \textit{\uppercase{L}ie \uppercase{G}roups, \uppercase{L}ie \uppercase{A}lgebras and \uppercase{s}ome of \uppercase{t}heir \uppercase{A}pplications}, Dover Publications, Inc.  436 (2005).



\bibitem {CeTa3} E. Celeghini and M. Tarlini,  Il Nuovo Cimento B \textbf{68}, 133 (1982).

\bibitem {Wei44} E. Weimar-Woods,  Rev. Math. Phys. \textbf{18}, 655 (2006).



\bibitem {Faw} R. J. B. Fawcett and A. J. Bracken,  J. Math. Phys. \textbf{29}, 1521 (2006).

\bibitem {Tal} J. D. Talman,  \textit{Special functions a group theoretical approach}, W. A. Benjamin, Inc. (1968).


\bibitem {MickN}J. Mickelsson and J. Niederle,   Commun. math. Phys.  \textbf{27}, 167 (1972).

\bibitem {Ball} A. Ballesteros, F. J. Herranz, O. Ragnisco and M. Santander,  \textit{Contractions, deformations and curvature} arXiv:0706.2208v1.

\bibitem {Segal} I. E. Segal,  Duke Math. J. \textbf{18}, 221 (1951).

\bibitem {Sal} E. J. Saletan,  J. Math. Phys. \textbf{2}, 1 (1961).


\bibitem {Wei1} E. Weimar-Woods,  J. Math. Phys. \textbf{32} (8), 2028 (1991).

\bibitem {Wei00} E. Weimar-Woods,  Rev. Math. Phys. \textbf{12},  1505 (2000).
\bibitem {Mon} M. de Montingny and J. Patera, J. Phys. A: Math. Gen. \textbf{24}, 525  (1991).


\bibitem {Conatser} C. W. Conatser, J. Math. Phys. \textbf{13}, 196 (1972).

\bibitem {Hud} P. L. Huddleston,  J. Math. Phys. \textbf{19} (8), 1645 (1978).
\bibitem {Nes} M. Nesterenko and R. Popovych,  J. Math. Phys. \textbf{47}, 123515 (2006).
\bibitem {Fla1} A. Fialowski and M. de Montigny,  J. Phys. A: Math. Gen. \textbf{38}, 6335 (2005).
\bibitem {Fla2} A. Fialowski and M. Penkava,  International Journal of Theoretical Physics, \textbf{47}, 531 (2008).



\bibitem {Mack} G. W. Mackey,  Proc. Nat. Acad. Sci. U.S \textbf{35}, 537 (1949).


\bibitem {Vil} N. Ja. Vilenkin,   \textit{Special Functions and the Theory of Group Representations} Amer. Math. Soc., Providence (1968).
\bibitem {Baruch} E. M. Baruch,  The classical Hankel transform in the Kirillov model, \textit{In preperation}

\bibitem{Izm1} A.A. Izmest'ev, G.S. Pogosyan, A.N. Sissakian and P. Winternitz, J. Phys. A: Math. Gen. \textbf{29}, 5940 (1996).
\bibitem{Izm2} A.A. Izmest'ev, G.S. Pogosyan, A.N. Sissakian and P. Winternitz, Int. J. Mod. Phys. A \textbf{12}, 53 (1997).
\bibitem{Kal} E.G. Kalnins, W. Miller Jr. and G.S. Pogosyan, J. Phys. A: Math. Gen. \textbf{32}, 4709 (1999).
\bibitem{Izm3} A.A. Izmest'ev, G.S. Pogosyan, A.N. Sissakian and P. Winternitz, J. Math. Phys. \textbf{40} (3), 1549  (1999).
\bibitem{Izm4}  A.A. Izmest'ev, G.S. Pogosyan, A.N. Sissakian and P. Winternitz, J. Phys. A: Math. Gen. \textbf{34}, 521  (2001).

\bibitem{Pog} G.S. Pogosyan, A.N. Sissakian and P. Winternitz, Phys. Part. Nuclei \textbf{33} (7), 235-276 (2002).


\bibitem {Mac} Saunders Mac Lane,   \textit{Categories for the Working Mathematician} Springer, 2nd edition (1998).

\bibitem {Cat} U. Cattaneo and W. Wreszinsky,  Commun. Math. Phys. \textbf{68}, 83 (1979).

\bibitem {CeTa2} E. Celeghini and M. Tarlini,  Il Nuovo Cimento B \textbf{65}, 172 (1981).



\bibitem {CW} U. Cattaneo and W. F. Wreszinski,  Rev. Math. Phys.  \textbf{11},  1179 (1999).




\bibitem{Pat} R. V. Moody and J. Patera, J. Phys. A: Math. Gen. \textbf{24}, 2227  (1991).

\bibitem {Wal} Nolan R. Wallach,  \textit{Real Reductive Groups I} Academic Press, Inc. (1988).  Nolan R. Wallach,  \textit{Real Reductive Groups II} Academic Press, Inc. (1992).

\end{thebibliography}
\end{document}